\documentstyle[11pt,aaspp4]{article}
\begin{document}
\newcommand{\lya}{Lyman~$\alpha$}
\newcommand{\lyb}{Lyman~$\beta$}
\newcommand{\za}{$z_{\rm abs}$}
\newcommand{\ze}{$z_{\rm em}$}
\newcommand{\cmtwo}{cm$^{-2}$}
\newcommand{\nhi}{$N$(H$^0$)}
\newcommand{\degpoint}{\mbox{$^\circ\mskip-7.0mu.\,$}}
\newcommand{\kms}{\,km~s$^{-1}$}      
\newcommand{\minpoint}{\mbox{$'\mskip-4.7mu.\mskip0.8mu$}}
\newcommand{\peryr}{\mbox{$\>\rm yr^{-1}$}}
\newcommand{\secpoint}{\mbox{$''\mskip-7.6mu.\,$}}
\newcommand{\sqdeg}{\mbox{${\rm deg}^2$}}
\newcommand{\squig}{\sim\!\!}
\newcommand{\subsun}{\mbox{$_{\twelvesy\odot}$}}
\newcommand{\et}{{\it et al.}~}

\def\ltsima{$\; \buildrel < \over \sim \;$}
\def\simlt{\lower.5ex\hbox{\ltsima}}
\def\gtsima{$\; \buildrel > \over \sim \;$}
\def\simgt{\lower.5ex\hbox{\gtsima}}
\def\arcs{$''~$}
\def\arcm{$'~$}
\def\erf{\mathop{\rm erf}}
\def\erfc{\mathop{\rm erfc}}
\title{LYMAN BREAK GALAXIES AT $Z \simgt 4$ AND THE EVOLUTION OF THE UV LUMINOSITY DENSITY AT HIGH REDSHIFT\altaffilmark{1}}
\author{\sc Charles C. Steidel\altaffilmark{2} and Kurt L. Adelberger}
\affil{Palomar Observatory, Caltech 105--24, Pasadena, CA 91125}
\author{\sc Mauro Giavalisco and Mark Dickinson}
\affil{Space Telescope Science Institute, 3700 San Martin Drive, Baltimore, MD 21218}
\author{\sc Max Pettini}
\affil{Royal Greenwich Observatory, Madingley Road, Cambridge CB3 0EZ, UK}

\altaffiltext{1}{Based on data obtained at the Palomar Observatory, the Cerro--Tololo
Inter-American Observatory, the William Herschel Telescope, and the W.M. Keck
Observatory. The W. M. Keck Observatory
is operated as a scientific partnership among the California Institute of Technology, the
University of California, and NASA, and was made possible by the generous financial
support of the W.M. Keck Foundation.
} 
\altaffiltext{2}{NSF Young Investigator}
\begin{abstract}
We present initial results of a survey for star-forming galaxies in the redshift range
$3.8 \simlt z \simlt 4.5$. 
This sample consists of
a photometric catalog of 244 galaxies culled from a total solid angle of 0.23 square
degrees to an apparent magnitude of $I_{\rm AB}=25.0$. Spectroscopic redshifts 
in the range $3.61 \le z \le 4.81$ have been obtained for 48 of these galaxies;
their median
redshift is $\langle z \rangle = 4.13$. 
Selecting these galaxies in a manner
entirely analogous to our large survey for Lyman break galaxies at smaller
redshift ($2.7 \simlt z \simlt 3.4$) allows a relatively clean
differential comparison between the populations and integrated luminosity density
at these two cosmic epochs. 
Over the same range of UV luminosity, the spectroscopic properties of
the galaxy samples at $z\sim 4$ and $z\sim 3$ are indistinguishable,
as are the luminosity function shapes and the total integrated
UV luminosity densities ($\rho_{UV}(z=3)/\rho_{UV}(z=4) = 1.1 \pm 0.3$).
We see no evidence at these bright magnitudes for the steep decline
in the star formation density inferred
from fainter photometric Lyman-break galaxies in the Hubble Deep Field (HDF).

The HDF provides the only existing data on Lyman-break galaxy number densities
at fainter magnitudes.  We have reanalyzed the $z\sim 3$ and $z\sim 4$ Lyman-break galaxies
in the HDF using our improved knowledge of the spectral energy distributions of these
galaxies, and we find, like previous authors, that faint Lyman-break galaxies appear
to be rarer at $z\sim 4$ than $z\sim 3$.  This might signal a large
change in the faint-end slope of the Lyman-break galaxy luminosity function between
redshifts $z\sim 3$ and $z\sim 4$, or, more likely, be due to
significant variance in the number  counts within the small volues probed
by the HDF at high redshifts ($\sim 160$ times smaller than the ground--based
surveys discussed here). 
If the true luminosity density at $z\sim 4$ is somewhat higher than
implied by the HDF, as our ground-based sample suggests, then
the emissivity of star formation as a function of redshift
would appear essentially constant for all $z >1$ once internally
consistent corrections for dust are made.  This suggests that there may be no obvious
peak in star formation activity, and that the onset of
substantial star formation in galaxies might occur at $z \simgt 4.5$.

\end{abstract}

\section{INTRODUCTION}

Within the last few years it has become possible to undertake large 
surveys of galaxies at very large redshifts ($z > 2$).  Simple 
photometric techniques keyed to the passage of the Lyman break 
through broad-band filters allow efficient selection of high 
redshift galaxy candidates (Steidel, Pettini \& Hamilton 1995;
Steidel \et 1996a,b;  Madau \et 1996) which can 
then be confirmed and studied thanks to the large spectroscopic 
throughput of the Low Resolution Imaging Spectrograph on the 
W.M. Keck 10m telescopes (Oke \et 1995).  It is now
feasible to study the large-scale distribution of star-forming
galaxies at high redshift (Steidel \et 1998; Giavalisco \et 1998; Adelberger \et 1998), 
and to obtain large enough samples of galaxies that accurate luminosity functions,
color distributions, and the like for $z \sim 3$ objects can be compiled (e.g., Dickinson
1998; Steidel \et 1998b).  The advantage of large and reasonably well defined samples
is that they allow direct comparisons to the predictions of galaxy
and structure formation models (e.g., Baugh \et 1998; Governato \et 1998; Katz, Hernquist,
\& Weinberg 1998;
Wechsler \et 1998; Somerville, Primack, \& Faber 1998; Coles \et 1998; Bagla 1998; Jing \& 
Suto 1998). 

Most of the work up to the present has concentrated on the redshift
regime $2.5 \simlt z \simlt 3.5$, for primarily practical reasons.  These
are the lowest redshifts at which the so-called ``Lyman-break''
technique can be applied using ground-based
photometry.  Also, at $z\sim 3$ the spectroscopic features that are most useful for
determining the redshifts of Lyman-break galaxies (LBGs) fall
within the wavelength range for which optical
spectrographs are most efficient, and in which the night sky background is minimal. As discussed
by Steidel \et 1998b, while conceptually it is straightforward
to extend the Lyman-break technique to higher redshift
(one simply uses a filter system that is shifted to longer wavelengths), 
the spectroscopic confirmation of photometrically selected objects becomes
far more difficult.

The Lyman-break technique has already been used in a very
powerful way in the Hubble Deep Field to estimate the star formation history of the Universe
in coarse redshift bins defined by galaxies' Lyman breaks passing through 
the F300W, F450W, and F606W filters (Madau \et 1996; Madau, Pozzetti, \& Dickinson 1998,
hereafter MPD).
It has now become a major industry to place any observation of high redshift
objects in the context of their implied contribution to the history of star
formation, using the Hubble Deep Field (HDF) observations as the
basis for comparison at high redshifts ($z \simgt 2$). 
Possibly one of the most important results from a series of papers by
Madau and others is the large apparent increase in ultraviolet
luminosity density between redshift bins centered at 
$\langle z \rangle \approx 4$ and $\langle z \rangle \approx 2.75$,
which suggests a rapid increase in the co-moving
volume-averaged rate of star formation over a rather short
interval of cosmic time.  This intriguing result appears to be consistent 
with what is observed for the space density of luminous, high
redshift QSOs (e.g.  Schmidt \et 1995, Kennefick 
\et 1995, Shaver \et 1998), and implies that
at $z\simgt 4$ one is entering the ``dark ages,'' the epoch when
star formation was first turning on in galaxies.  Some other studies
using photometric redshifts in the HDF, however, did not find evidence
for such a clear change in the UV luminosity density over this redshift
range (e.g. Sawicki \et 1997; Pascarelle \et 1999).

There are several reasons to be concerned about 
results based solely on the Hubble Deep Field, however.  First, the HDF, while
clearly the highest quality image of the sky ever obtained, is after all only
a very small piece of sky ($\simeq 5$ arcmin$^{2}$), and samples a relatively small volume at any redshift.
Since observations of the Lyman-break galaxies at $z \sim 3$
in ground-based surveys have shown that high redshift (luminous) star-forming
galaxies are strongly clustered (Steidel \et 1998a, Giavalisco \et 1998, Adelberger
\et 1998), one might be concerned about sample variance associated with a relatively small
volume.  
Moreover, even if the HDF provided a fair sample of
the universe, the redshift distributions of F300W and F450W dropouts
are not well known empirically. As a result, the effective volumes
used by Madau \et to calculate the star-formation densities at $\langle z\rangle =2.75$
and $\langle z\rangle=4$ were based upon models of the
spectral-energy distributions and Lyman-continuum opacities of galaxies, and
not on spectroscopic redshifts. We have found, in our large survey at
$z\sim 3$, that the effective redshift selection function imposed by
a particular set of photometric selection criteria is not a simple
``boxcar'' as assumed by Madau \et\,
Objects near the assumed boundaries
of the $N(z)$ function are under--represented due to photometric errors, and, more
importantly, 
to the fact that there are substantial variations in the spectra of galaxies at a given
redshift. These variations are due, among other things, to the stochastic nature of the 
line blanketing in the Lyman $\alpha$ forest, to the amount
of intrinsic reddening by dust, and to whether Lyman $\alpha$ is in emission or
absorption. 
While such subtleties may not seem important for the rather
crude luminosity densities estimated from photometric redshifts,
they undoubtedly have at least some effect on the implied luminosity
densities, and the magnitude of this effect is difficult to estimate without spectroscopic
redshifts.

In view of the importance of confirming the decline in the far-UV
luminosity density between $z \sim 3$ and $z\sim 4$, and the fact
that we now have a substantial sample of galaxies in the redshift
range $2.6 \simlt z \simlt 3.4$ from our relatively wide-angle
ground-based surveys, we have undertaken to compile wide-field
photometry and spectroscopic confirmation of Lyman-break galaxies in a redshift
interval $3.8 \simlt z \simlt 4.5$.

Although comparatively low spatial resolution and significantly brighter backgrounds make
it difficult to reach the depth of the HDF from the ground, it is feasible to    
to survey much larger fields;  our present survey aims
to exploit this advantage.
We have attempted to reach depths for $z\sim 4$ LBGs that are
comparable to our existing survey at $z\sim 3$, over a region $\sim 160$ 
times larger than the HDF.  
The $z\sim3$ survey has been based entirely on the custom $U_nG{\cal R}$
filter system (described in detail in Steidel \& Hamilton 1993), with
effective wavelengths of 3650, 4750, and 6930 \AA, respectively. 
It turns out to be remarkably efficient, for the purposes of
selecting Lyman-break galaxies at $z \simgt 4$, to add one passband
($I$, with an effective wavelength of 8100 \AA)
to our existing imaging data, and to select $G$-band dropouts using
the $G {\cal R}I$ system in a manner entirely analogous to the way
we have used $U_nG{\cal R}$ at $z \sim 3$. 

In this paper, we present the initial results of our galaxy survey
with an expected median redshift of $\langle z \rangle = 4.2$, to
demonstrate the feasibility of extending spectroscopic Lyman-break
galaxy surveys to higher redshifts, and to make a preliminary estimate
of the star formation luminosity density for comparison to the analogous
value at $z \sim 3$, all based on data that are independent of the
Hubble Deep Field.  

\section{PHOTOMETRIC SELECTION CRITERIA}

The photometric selection criteria, as for our $z \sim 3$ sample, are
based upon a combination of the expectations from modeling the
spectral energy distributions of star-forming objects at high
redshift, combined with practical considerations such as allowing
for photometric errors, and steering away from regions in the
$G{\cal R}I$ color--color plane that are obviously contaminated with
objects that are {\it not} at very high redshift (i.e., ``interlopers''). 
Our intention
initially was to err on the side of caution and obtain redshifts
for objects with a rather broad range of colors, since
we did not know {\it a priori} where in the $G{\cal R}I$ color--color plane the population of
$z \sim 4$ objects would lie.  

We defined the initial color selection
criteria with the observed range of intrinsic $z \sim 3$ galaxy
colors in mind. As will be discussed elsewhere (Adelberger \et 1999), the
colors of the galaxies in our $z \sim3$ sample are consistent
with standard Bruzual and Charlot (1996) models of continuous
star formation, altered by the statistical opacity of the IGM
(following Madau 1995) and a component of optically thick H~I
in the galaxy itself\footnote{This component has not generally been included 
in past work}, and reddened by applying a version
of the starburst galaxy obscuration relation of Calzetti (1997) extrapolated
to wavelengths shorter than 1200 \AA\footnote{In the extrapolation we
assumed that the relation continues to rise from 1200 to 912 \AA, as is the 
case for the Galactic extinction curve (Snow, Allen, \& Polidan 1990). However,
we note that the exact form of the extrapolation has only a minor effect on the
predicted $U_n-G$ color of the models.}
The range of UV
colors observed in the $z \sim 3$ sample is well represented
by such models with $E(B-V)$ ranging from zero to $\sim$0.3 magnitudes,
with a median color (which can be thought of as the spectrum
of the ``typical'' Lyman-break galaxy in the $z \sim 3$ sample)
corresponding to $E(B-V)=0.15$ for the adopted
reddening curve (note that this corresponds to an extinction at
rest-frame $\sim 1700$ \AA\ [the observed ${\cal R}$ band] of about a factor of 4). 
All magnitudes and colors used in this paper are reported on the
``AB'' system (Oke \& Gunn 1983). 

Our $G{\cal R}I$ photometric selection criteria were designed to select galaxies
at $z\sim 4$ with a range of intrinsic SEDs similar to what is observed in our $U_nG{\cal R}$
sample at $z\sim 3$.
In Figure 1a, we have plotted the tracks
in the $G{\cal R}I$ color--color plane for the model spectra  which closely match
the observed galaxies at $z \sim 3$; the same model galaxies are plotted in Figure 1b
to show their location relative to the color selection window used for the
$z \sim 3$ sample.
There are two different color selection windows indicated
in Figure 1a. The shaded region is the one actually used to select objects
for spectroscopic follow-up, which was intended to be relatively broad 
in order to explore the color--color space (and the effects of photometric
scatter) somewhat; the bold line encompasses the region which
we will use as our primary selection window for comparison with
the $z \sim 3$ data. The latter was adopted because it would result in
comparable effective volumes near $\langle z \rangle = 4.2$
and $\langle z \rangle = 3.05$ in our two surveys if galaxies
had the same range of intrinsic colors at these redshifts,
and also because (as we shall see below) it
turns out that one can eliminate a substantial fraction of the
interlopers by keeping the ${\cal R}-I$ colors relatively blue. 
This selection window is defined by
$$G-{\cal R}\geq 2.0,\quad G-{\cal R}\geq 2({\cal R}-I) + 1.5,\quad {\cal R}-I\leq 0.6.$$

The ``median color'' galaxy model (the middle of the three color-color tracks shown
in Figure 1a)
enters the primary selection window at $z \sim 3.9$, and exits again 
at $z \sim 4.5$, so that this would be the {\it a priori} expected 
redshift range of a galaxy sample similar to the one observed
at $z \sim 3$. This figure also shows that bluer (i.e., less reddened) galaxies are expected to be biased somewhat
toward the higher redshifts, and redder (i.e., more reddened) ones toward the lower redshifts;
this is the main effect that makes the realized $N(z)$ function different
from a ``boxcar'' 
(see, e.g., Steidel \et 1998b, and Figure 4). In Figure 1a we have also plotted the
color tracks for 
a template elliptical galaxy, with only k-corrections applied
(i.e., no spectral evolution). Note that the locus of unevolved early type galaxies
comes very close to the selection
window for the $z \sim 4$ galaxies for $z \sim 0.5-1$.  A combination of
photometric errors and intrinsic variations in galaxy spectral-energy distributions
can scatter some early-type galaxies at these
redshifts into our selection window.
It turns out that such objects are the sole source of ``interlopers'' for
the $z \simgt 4$ sample, and
we suggest below ways that their contribution can be minimized. Note also
that no contamination by stars is expected (unlike in the
$z \sim 3$ sample---see Figure 1b).

\section{OBSERVATIONS}

\subsection{Imaging}

With the exception of DSF1550, all of the fields included here have been
part of our extensive survey for $z \sim 3$ galaxies; the field
centers are given in Table 1.
The imaging 
data were obtained at the William Herschel telescope (3C 324 and B2 0902+34), Palomar
200-inch Hale telescope (CDFa, CDFb, DSF2237a, DSF2237b, SSA22a, SSA22b, and B2 0902+34), and the CTIO 4m telescope
(DSF1550+08) during the interval 1996-98. All of the imaging data and the photometric
methods employed will be presented in detail elsewhere; 
in brief, the photometry was performed in a manner identical to that used for the $ z\sim 3$
galaxy searches, which is described (for example) in Steidel, Pettini, \& Hamilton (1995), with
the exception that we have used a modified version of the FOCAS (Valdes 1982) image
detection and analysis routines that one of us (KLA) has optimized for our purposes. 

For fields in which existing $U_nG{\cal R}$ catalogs were already on hand, we simply
registered the $I$ image onto the same coordinate system and obtained the $I$ magnitudes
and ${\cal R}-I$ colors through the same matched apertures. We have performed
our object detection on the ${\cal R}$ band images because they are in most cases
significantly deeper than in the $I$ band. However, the selection of candidate $z \sim 4$
galaxies has been made using the $I$ band total magnitudes. (This procedure is unlikely
to present any significant biases, given that the expected colors are quite blue
in ${\cal R}-I$, and our ${\cal R}$ images typically reach $\sim 0.5$ magnitudes deeper than
the $I$ images). The typical depths of our images, in $\sim 1$\arcs\
seeing, are 29.3, 28.5, and 28.0 magnitudes per square arc second for
$G$, ${\cal R}$, and $I$, respectively, within 
a 1\arcs\ aperture (1 $\sigma$). Thus, an object with $I=25$ is approximately
a $5-10\sigma$ detection in the $I$ band, with considerable dynamic range available
to measure colors in the $G$ and ${\cal R}$ bands. 

We have chosen to limit the object catalogs to $I\leq 25.0$ in order to provide
a relatively high level of photometric completeness and ensure that there would be sufficient
dynamic range to detect breaks in the $G-{\cal R}$ colors.  We will discuss
completeness issues in \S 4 below.

The photometric selection criteria summarized in Figure 1a (shaded region) were
used to isolate candidates for $z \simgt 4$ objects. All such candidates were
examined visually in order to remove objects that were clearly spurious (these usually
were found near very bright stars). A summary of the number of remaining candidates
in each field is given in Table 1.  A composite 2-color diagram from the 10 fields
included in this paper is shown in Figure 2.

\subsection{Spectroscopy}

All of the galaxy spectra were obtained using the Low Resolution Imaging Spectrograph 
(Oke \et 1995) on the Keck II 10m telescope, between 1997 March and 1998 October.
For the observing runs in 1997, we generally included several slits targeting $z \sim 4$
candidates on masks designed primarily for our $z \sim 3$ LBG survey; as a consequence,
most of these spectra were obtained using the 300 line/mm grating blazed at 5000 \AA\ 
and with a grating tilt optimized for the 4000--7000 \AA\ range. A more efficient 
configuration, with better sensitivity in the crucial 6000-7500 \AA\ range, was to use
a 150 line/mm grating blazed at 7500\AA; this was used in the 1998 observing runs and increased the
spectroscopic success rate by about a factor of 2--3 for both $z\sim 4$ galaxies
and lower redshift interlopers (e.g., a single slit mask, with 16 candidate
objects, yielded 7 successful redshifts in the $3.9 \le z \le 4.5$ and one $z=0.96$
interloper in the DSF1550 field in 1998 May---the remaining 8 candidates had
inadequate S/N for identification). Slit widths on the masks were
either 1\secpoint0 or 1\secpoint4, resulting in spectral resolution ranging
from 10-12\AA\ for the 300 line configuration and 20-25\AA\ for the
150 line configuration.  Typical total exposure times per mask
were 2 hours, usually broken into individual exposures of 1200 or 1800s, with small
dithers along the slitlets between exposures in order to sample different parts
of the detector and to allow the option of various schemes for the removal of fringes at
redder wavelengths. The data were reduced using a suite of custom IRAF scripts.

Examples of $z\sim 4$ galaxy spectra are shown in Figure 3. 
The onset of strong Lyman $\alpha$ forest blanketing is very apparent
in the spectra of these galaxies.
As in the $z \sim 3$ sample, there is a wide variety of spectroscopic properties, ranging
from Lyman $\alpha$ in emission with rest-frame equivalent widths up to $\sim 80$ \AA, to
objects with very strong Lyman $\alpha$ absorption and accompanying strong lines of
various interstellar transitions of low ions, to objects
that have Lyman $\alpha$ in both absorption and in emission (see, e.g., the spectrum
of HDF G4 in Figure 3). There is an obvious spectroscopic bias against objects
that do not have strong emission lines, but even with this bias about half of our
successful redshifts are based purely on absorption features.  The substantial fraction
of absorption-dominated spectra in our $z\sim 4$ sample may be due in part
to the large intrinsic luminosities of the galaxies; we see some evidence
in our $z\sim 3$ sample, which reaches fainter intrinsic luminosities, that
brighter objects tend to have absorption-dominated spectra. These effects will
will be quantified elsewhere.

Unsuccessful redshifts were invariably
due to lack of adequate S/N. This is a more serious problem at
$z\sim 4$ than at $z\sim 3$, largely because the spectral
features used to identify redshifts are moved from
4500--6000 \AA\ to 6500--7500 \AA, where the sky is 1.5--2 magnitudes
brighter.  As a result our success rate for identifying redshifts
is $\sim$30--50\% at $z\sim 4$, compared to $\simgt 80$\% at $z\sim 3$.
Objects with strong Lyman $\alpha$ emission lines are clearly
less likely to fall into the ``unsuccessful'' category, as the continuum S/N is
much less important than for the absorption line objects, and for this reason it
is likely that most of the spectra which remain to be identified do not have
Lyman $\alpha$ emission with large equivalent width.  In
the range of continuum luminosity that we are currently probing, even very
sensitive narrow-band searches would turn up only a relatively small fraction of
the galaxies (cf. Hu \et 1998). 

The magnitudes, colors, and redshifts of all 48 of the spectroscopically confirmed $z\simgt 4$ galaxies
are summarized in Table 2. Also summarized are the same quantities for the galaxies
identified as interlopers; all 73 of the objects with redshifts are indicated with
shaded ``dots'' on Figure 2. From Figure 2, it is clear that most of the
objects in the selection window with red ${\cal R}-I$ colors are
interlopers. As mentioned above, by tightening the color selection window
to include only objects having ${\cal R}-I \le 0.6$, one immediately eliminates
11 of the 25 interlopers, most of which have $z\sim 1$, as expected from Figure
1a; only 2 out of 48 of the bona fide high redshift galaxies are excluded\footnote{One of these,
CDFa-G1 at $z=4.812$, is the most intrinsically luminous Lyman break galaxy in our
entire sample of more than 700 at $z >2$. The relatively red ${\cal R}-I$ color
in this case is due to significant blanketing of the ${\cal R}$ passband by the
Lyman alpha forest.}.  
In addition, applying the requirement that the galaxies must
have $I \ge 23.0$ (this will tend to screen out $z \sim 0.5$ early type galaxies,
which will typically have $I \sim 21-22$) eliminates an additional 2 interlopers
in the spectroscopic sample, and has no effect on the true high redshift sample
for objects with redshifts. 
Thus, with this adjustment of the color selection criteria (which 
one would have chosen in any case in order to observe the same
types of galaxies seen in the $z \sim 3$ sample, as demonstrated in Figure 1), the
contamination of the spectroscopic sample can be reduced from 25/73 to 12/60,
with the loss of only 2 true high redshift galaxies. As can be seen from Table 1,
applying the new color cut reduces the number of candidates from 244 to 207. 
We regard this sample of 207 candidates as our primary photometric sample\footnote{
The effect on the net surface density of candidates after correction for interlopers
is barely affected by this cut; the contamination-corrected number of candidates
is $244(48/73)=160$ without the cut, and $207(46/58)=164$, or a difference of much less
than 1 $\sigma$)}, and the estimated contamination by interlopers is $\sim 20\%$.

\section{DIFFERENTIAL COMPARISONS: $z \sim4$ VERSUS $z \sim 3$}

\subsection{Correcting for Incompleteness}

The observed surface density of $z \sim 4$ galaxies to a limit of $I=25.0$,
$\Sigma_4(25) = 0.20 \pm 0.02$ arcmin$^{-2}$, is
significantly smaller than the observed surface density of $z \sim 3$ galaxies to ${\cal R}=25.0$, 
$\Sigma_3(25) = 0.68\pm0.03$ arcmin$^{-2}$. The $I$ band samples the far--UV continuum of $z\sim 4$
galaxies at very similar rest wavelengths as the ${\cal R}$ band at $ z\sim 3$ (by design),
so that k-corrections between the two should be negligible; however, obviously
one is sampling different parts of the luminosity function because of the larger luminosity
distance for the $z \sim 4$ sample. Our approach for comparing the two samples will be to
truncate the $z\sim 3$ sample at an apparent ${\cal R}$ magnitude that corresponds to 
an absolute far--UV flux density equivalent to $I=25.0$ at $z=4.13$ (the median redshift
of the $z \sim 4$ sample). This apparent magnitude is cosmology dependent: 
${\cal R}=24.51$ for Einstein-de Sitter,  ${\cal R}=24.31$ for $\Omega_0=0.2$ open,
and ${\cal R}=24.45$ for $\Omega_0=0.3$, $\Omega_{\Lambda}=0.7$. 

The first step is to correct the samples for 
contamination by interlopers. The $z \sim 4$ surface density was
statistically corrected for contamination based on the spectroscopic results, as described
above.
At $z \sim 3$, the only known interlopers (other than QSOs, which we have also
removed from the sample)
are Galactic stars, which we have found empirically to have a surface
density to ${\cal R}=24.0$ along the sightlines we sample
of $\Sigma_{\rm stars} = 0.06$ arc min$^{-2}$ (and
essentially zero at fainter apparent magnitudes). Thus, both samples have comparable
contamination by interlopers, and both interloper populations (stars and
early type galaxies) tend to populate the bright end of the apparent magnitude
distribution; all corrections for contamination have been made as a function
of apparent magnitude. 

Once interlopers have been removed from the samples, we can begin
to estimate the comoving luminosity density at $z\sim 3$ and
$z\sim 4$ from the observed surface density of $U_n$ and $G$ dropouts.
The main complication is that some fraction of the galaxies brighter
than our magnitude limits at these redshifts are undoubtedly missing
from our sample because 
of photometric errors, blending with foreground objects,
and---more seriously---because they may have intrinsic colors which
do not match our selection criteria.  In order to investigate
this incompleteness quantitatively, we have run Monte-Carlo simulations
in which large numbers of artificial objects with a range of
colors and magnitudes are added to our real images, and then recovered using
the same methods employed for the real photometry.  The complete results
of these simulations will be presented in Adelberger \et (1999, A99);
but for the present they are useful mainly because they yield
an estimate of the effective volume of our surveys as a function
of apparent magnitude, 
$$V_{\rm eff}(m) \equiv \int dz\, p(m,z) dV/dz,$$
where $p(m,z)$ is the probability a galaxy of apparent magnitude
$m$ (in ${\cal R}$ at $z\sim 3$ and $I$ at $z\sim 4$) at redshift $z$
will be detected in our images and appear to match our photometric
criteria, and $dz\, dV/dz$ is the comoving volume per arcmin$^2$
at redshift $z$.  As shown in A99, dividing the observed
surface density of high redshift candidates by $V_{\rm eff}(m)$
defined in this way compensates for our various selection biases
and incompletenesses, and produces a maximum likelihood estimate
of the comoving number density of galaxies with magnitude $m$
at the observed redshifts.  The galaxy number density estimated
with this procedure is corrected not only for the usual
``detection incompleteness'' due to the declining probability of
detecting a galaxy in our images at the faintest magnitudes
(caused by both photometric errors and problems related to blending
with foreground objects\footnote{The completeness correction for even
the brightest magnitude bin is $\sim 30$\%, due mostly to blending
effects.}), but
also for ``color incompleteness'' due to detected galaxies having
measured colors that may erroneously place them outside of our color-selection window,
and even to some extent for the ``template incompleteness'' which
would arise if some fraction of high-redshift galaxies had true 
colors that lay outside of our color-selection window.
This last correction is possible because our $U_nG{\cal R}$ sample (for example)
could contain galaxies at $z\sim 2.5$ much redder than the typical color
we assume, and galaxies at $z\sim 3.5$ much bluer, and we can therefore
use data from these redshifts to estimate the fraction of galaxies whose colors will not match
our selection criteria at $z\simeq 3$, where most of our sample lies.
We will return to this below.

The effective volumes calculated from our Monte-Carlo simulations are
provided, as a function of apparent magnitude, in Tables 3 and 4.  
The incompleteness corrections implied by these effective volumes
are significant, but because the corrections are comparable (to within $\sim$ 30\%)
for the $z\sim 3$ and $z\sim 4$ samples, they do not strongly affect our
estimate of the change in luminosity density over this range of redshift.
Very similar relative volumes at $z\sim 3$ and $z\sim 4$
would result, for example, from simply assuming a boxcar selection
function with half-width given by the
standard deviation (see Figure 4) of the redshifts in each sample.
The incompleteness corrections are important when we
attempt to derive absolute UV luminosity densities at $z\sim 3$
and $z\sim 4$, but this calculation is subject to 
other, probably larger, uncertainties, as we discuss below.

\subsection{Color and Luminosity Distributions}

From these effective volumes it is straightforward to compute
luminosity functions from the observed surface density of $U_n$
and $G$ dropouts. 
It is of primary interest to compare the luminosity functions, and
the integral luminosity densities, in the two redshift
intervals. The most recent compilation from the Hubble Deep Field data is
that of Madau, Pozzetti, \& Dickinson (1998, MPD), which suggests a factor of
$\sim 2.5$ decline in the luminosity density between $z \sim 2.75$ and $z \sim 4$
(the original Madau \et 1996 paper estimated the same decline at a factor
of $\sim 4$).  In contrast we see little evidence in our ground-based sample
for luminosity density evolution from $z\sim 4$ to $z\sim 3$.
Figure 5 compares the $U_n$- and $G$-dropout luminosity
functions for an Einstein-de Sitter
cosmology. 
The agreement in both the shape and normalization of the bright end
of the luminosity functions at $z \sim 3$ and $z \sim 4$ is quite striking, and
depends only mildly on cosmology 
(the other 2 cosmologies considered make the $z \sim 4$ luminosity
function slightly brighter relative to that at $z \sim 3$).  To construct this
plot from the comoving number density of objects with apparent magnitude
$m_{\rm AB}$, which we estimated above, we used the relationship
$L_\nu = 10^{-0.4(48.60+m_{\rm AB})} 4\pi d_L^2 / (1+z)$ 
and adopted (for simplicity) $\Omega_m=1$ luminosity distances of
$d_L = 3.8\times 10^{28} h^{-1}$ cm for all objects in our $z\sim 3$
sample, and $d_L = 5.3 \times 10^{28} h^{-1}$ cm for all objects in our
$z\sim 4$ sample.  These luminosity distances are appropriate for
objects at the median observed redshifts of $\langle z\rangle =3.04$ and $\langle z\rangle=4.13$.
The corresponding luminosity distances for $\Omega_m=0.2$ open and
$\Omega_m=0.3$ flat are $d_L = 5.8, 5.6 \times 10^{28} h^{-1}$ cm 
at $\langle z\rangle =3.04$ and $d_L = 9.0, 8.1 \times 10^{28} h^{-1}$ cm 
at $\langle z\rangle =4.13$.

The results of integrating the luminosity distributions over the same range
of intrinsic luminosity 
(the smallest luminosity corresponds to the faintest objects in the $z \sim 4$ sample,
which is $I=25.0$, or $M_{AB}(1700\AA)=-21$ for the 
Einstein-de Sitter model $H_0=50$ \kms Mpc$^{-1}$ [see Figure 5]), as well as the ratios
of the luminosity densities in the two redshift intervals, are given in Table 5. 
In assigning errors to these luminosity densities, we have attempted to
account for Poisson counting statistics, uncertainties in the contamination corrections,
and for systematic uncertainties
in the effective volumes discussed above. 

For all cosmologies we consider, the observed UV-luminosity density integrated
to the luminosity equivalent of $I_{AB}=25.0$ at $z=4.13$      
shows no significant change between
$\langle z \rangle = 4.13$ and $\langle z \rangle =3.04$,
in apparent conflict with several analyses based upon the HDF alone (e.g.
Madau \et (1996), MPD, Pozzetti \et 1998).  We will return to this shortly.

A useful by-product of the estimated $V_{\rm eff}(m)$ from A99 is the
incompleteness-corrected distribution of intrinsic galaxy spectral shapes in our $z\sim 3$ sample,
shown in Fig. 6 (see A99 for details of how this plot was constructed).  
For convenience we have chosen to parameterize the range of LBG spectral
shapes by assuming that they arise due to varying amounts of reddening
by dust 
imposed on otherwise identical model galaxies (see \S 2).
This is obviously not an exact description of the physical situation,
as (for example) the galaxies with negative $E(B-V)$ in Fig. 6 show\footnote{
In almost all cases, galaxies with negative values of E(B-V) actually have continuum
shapes consistent with E(B-V)$=0$, but they  
have unusually
strong Lyman $\alpha$ emission (observed equivalent widths up to 300 \AA), which can affect the measured $G-{\cal R}$
color by a few tenths of a magnitude at the extreme, until $z \sim 3.35$, at
which point the Lyman $\alpha$ line moves out of the $G$ band. For the present purposes,
the incidence of such negative values of E(B-V) simply indicates the relative numbers
of such strong emission line galaxies. In the future, all galaxy colors will be corrected
for Lyman $\alpha$ contamination. In general, however, the effect is small.},and
in fact some of the range of observed $G-{\cal R}$ colors at $z\sim 3$
can be traced to fluctuations in the strength of Lyman $\alpha$
emission or absorption, and some is undoubtedly due to fluctuations
in intergalactic absorption and in star-formation histories among the
galaxies in our sample.  Nevertheless variable dust absorption is
plausibly the dominant factor (cf. Pettini \et 1998, Dickinson 1998, Meurer \et 1997, Calzetti
\& Heckman 1998), 
and for this reason we have chosen
to quantify deviations from the mean observed colors with $E(B-V)$.
For our present purposes the physical cause of the spread in measured colors
is irrelevant, however, and all that matters is that the family of
spectral shapes implicit in this parameterization
can reasonably approximate the entire range of far--UV spectral shapes in our $z\sim 3$ sample.  

An interesting aspect of Figure 6 is the clear dearth of objects
with $E(B-V) \simgt 0.3$.  Our sample is slightly biased against
reddened objects, but even after correcting for these selection
effects the comoving number density of galaxies with $E(B-V)\sim 0.1$
is more than ten times higher than that of galaxies with $E(B-V)\sim 0.4$.
We see no galaxies with $E(B-V)>0.5$, though these galaxies---if they existed---would
match our color criteria over the redshift range $2.5\simlt z\simlt 2.7$.\footnote{Although
a galaxy
with $E(B-V)\sim 0.5$ would have the right colors to be
included in our sample at $2.5\simlt z\simlt 2.7$,
the implied extinction (assuming the reddening were due to dust with the Calzetti (1997)
reddening curve)
would be 5.2 magnitudes, and these galaxies might
be missing from our sample simply because they are too faint.
However, we see no evidence in our sample that redder galaxies are systematically
fainter.  For example, the mean ${\cal R}$ magnitude of the 4 galaxies in our sample
with $0.4 < E(B-V) < 0.5$ is 24.6, somewhat brighter than average.}
Overall the histogram of $E(B-V)$ suggests that few galaxies at $z\sim 3$
are so red (or so blue) that they would be completely absent
in our $U_nG{\cal R}$ sample. The range of inferred E(B-V) is very similar
to the range observed in nearby starburst galaxies (Meurer \et 1995, 1997).

\subsection{The Hubble Deep Field}

With this distribution of $E(B-V)$ in hand, it is possible to predict the redshift
distributions of high-redshift color-selected samples with some accuracy.
Figure 4 shows a comparison of the observed redshift distributions
in our $U_nG{\cal R}$ and $G{\cal R}I$ samples with the redshift distributions
that would be expected, given the adopted color selection criteria, if 
spectral shapes were drawn randomly
from this histogram and the luminosity function were constant
within each redshift interval.  The agreement between the model predictions 
and the data,
for both the color distribution as a function of redshift, and the
overall redshift distribution, is reasonably good.  
This raises the hope that we might be able to get improved estimates of the
apparently discrepant luminosity densities implied by F300W and
F450W dropouts in the Hubble Deep Field if we used our distribution
of $E(B-V)$ to estimate the effective volumes for the HDF color-selection
criteria.  Figure 7 shows the redshift distributions we would predict
from our $E(B-V)$ distribution for galaxies selected according to the
criteria 
$$U_{300}-B_{450} \geq B_{450}-V_{606} + 1,\quad U_{300}-B_{450}\geq 1.6,\quad B_{450}-V_{606}\leq 1.2,\quad V_{606}\leq 27$$
and
$$B_{450}-V_{606}>1.5,\quad B_{450}-V_{606} > 1.7(V_{606}-I_{814})+0.7$$
$$B_{450}-V_{606} < 3.5(V_{606}-I_{814})+1.5,\quad V_{606}-I_{814}<1.5,\quad V_{606}<27.7$$
for the $U_{300}$ and $B_{450}$ dropouts, respectively
(cf. Dickinson (1998), Madau \et (1996)).  It has been assumed by Madau \et (1996), MPD,
Dickinson 1998, 
and Pozzetti \et 1998 
that these two selection criteria uniformly probe the redshift intervals
$2.0 < z < 3.5$ and $3.5 < z < 4.5$.  Our estimated redshift distributions are
broadly in agreement with this, although the effective volumes we find, 
380 (200) $h^{-3}$ comoving Mpc$^3$ arcmin$^{-2}$ for $\Omega=1$, are
13\% (19\%) smaller at $z\sim 3\,(4)$.  Unlike our effective volumes
for the ground-based samples, these effective volumes do not
include photometric errors; they account only for what
we have called ``template incompleteness'' above---i.e., for the fraction of galaxies
that would meet the HDF color-selection criteria in the absence of
photometric errors, if all galaxies had intrinsic spectral shapes
drawn from the distribution of A99 (see Figure 6 and the related discussion).  

Figure 8 shows our estimate of the faint end of the luminosity functions
at $z\sim 3$ and $z\sim 4$ derived from these effective volumes and
the observed surface density of $U_{300}$ and $B_{450}$ dropouts in the
HDF.  For the $U_{300}$ sample we used the approximation
${\cal R}\simeq (V_{606}+I_{814})/2$ and shifted all magnitudes 0.25 fainter to
account for its smaller mean redshift ($\langle z\rangle\sim 2.6$) compared to our $U_nG{\cal R}$ sample.
The number of $U_{300}$ dropouts in the HDF was taken from the catalog
of Dickinson (1998);
the $B_{450}$ dropouts are from table 3 of Madau \et (1996).  

The $U_{300}$ dropout points fall rather nicely along the Schechter (1976) function
with faint-end slope $\alpha\sim -1.6$ defined by the ground-based sample alone,
but this is likely at least partly fortuitous:  removing the $U_{300}-B_{450}>1.6$
criterion from our $U_{300}$ dropout definition would have changed our estimated effective
volume by only 14\%, but would have increased the number of $U_{300}$ dropouts
by about 45\% and significantly altered the luminosity function shape.  The large number of
dropouts excluded by the $U_{300}-B_{450}>1.6$ criterion have colors similar to
those expected for galaxies at $z\simeq 2$, suggesting that a significant
fraction of the $U_{300}$ dropouts in the HDF may be associated with a single
galaxy over-density at $z\sim 2$.  This would not be completely unexpected;
even in one of our much larger ground-based fields, ``SSA22a,'' almost half of
the $U_n$ dropouts lie within the narrow redshift interval $3.07<z<3.11$
(Steidel \et 1998a).
Regardless of the exact cause of these changes, it appears
that seemingly minor adjustments to color-selection criteria in the HDF
can have serious effects on the normalization and shape of derived luminosity
functions, and this illustrates that caution is required in interpreting the results
from a single small field as universal. 
Measuring consistent luminosity functions
in several fields spread over the sky would provide some assurance that the derived
luminosity densities may be close to their universal values, but currently
this is possible only at the relatively shallow depths reached in ground-based samples.

Nevertheless, we have fit a Schechter (1976) function to the combined HDF + ground-based
luminosity function shown in Figure 8, after excluding the faintest HDF point
which may suffer from incompleteness, and we find, with formal 68.3\% confidence,
that $m_*=24.48\pm0.15$ and $\alpha=-1.60\pm0.13$.  The fit is determined
largely by the ground-based data, and so our confidence in it is not entirely
eroded by our reservations about the HDF.  This estimate of the faint-end slope
$\alpha$ is considerably steeper than previous estimates (e.g. Dickinson 1998,
Pozzetti \et 1998)\footnote{The difference between the new luminosity function and that
presented by Dickinson (1998) is due to slightly modified selection criteria
in the HDF and (especially) to improved
estimates of incompleteness in the ground-based sample.}.

At $z\sim 4$ it is more difficult to gauge the consistency between the ground-based data
and the HDF, because there is very little luminosity overlap between the
two samples; only a single $B_{450}$ dropout in the HDF would have been bright enough
to be included in our ground-based sample.  A naive fit to the points in
the bottom panel of Figure 8 would
favor a faint-end slope of $\alpha\sim -1$, but because of our concerns about
sample variance in the HDF, we have chosen to plot (instead of an actual
fit) our $z\sim 3$
luminosity function redshifted to $z\sim 4$ in an $\Omega=1$ cosmology (i.e., with
$m^{\ast}$ shifted by the relative distance modulus between $z=3.04$ and $z=4.13$), and
multiplied by 0.8 (cf. Table 5).  
This matches our $z \sim 4$ data well at the bright end, but over-predicts the
number of fainter HDF $B_{450}$-dropouts by about a factor of two, and the integrated luminosity
density by about a factor of 2.5 (see the discussion in \S 5.1).
A combination of deeper ground-based data and the
HDF-South should help decide whether the disagreement of the HDF $B_{450}$ dropouts
at the faint end truly signals a different luminosity function shape at $z\sim 4$, or is
a further indication that the HDF does not provide a fair sample of the universe.

\subsection{The Evolution of the UV Luminosity Density}

Figure 8 contains most of the available information on the star-formation density
at $z\sim 3$ and $z\sim 4$, and we have used it as a basis for re-evaluating the
star-formation history of the universe derived previously by many authors,
most famously Madau \et (1996).  Our result is shown in the top panel
of Figure 9.  For consistency we estimated
the star-formation density of the universe at all redshifts by integrating
luminosity functions down to $0.1 L^*$, which corresponds roughly to
${\cal R}=27$, our faintest data point, at $z\sim 3$.  As discussed above,
the luminosity functions at $z\sim 3$ and (especially) $z\sim 4$ have not
been convincingly measured down to $0.1 L^*$, and this is one of the
many large uncertainties that underlie a diagram such as Figure 9.
At $z\sim 3$ we adopted the $m_*=24.48$, $\alpha=-1.60$ Schechter-function fit
derived above for the luminosity function.  
At $z\sim 4$ we used the luminosity function form shown in the bottom panel of
Figure 8; as discussed
above, this luminosity function agrees well with the $z\sim 4$ data at the
bright end, but is apparently inconsistent at the faint end
with the small number (13) of $B_{450}$ dropouts in the HDF.
The estimate of the star-formation density at $z\sim 1$ relies upon luminosity
functions at lower redshifts in a similar way:
because the high-redshift CFRS data do not extend fainter than 
$L^*$, the faint-end slope from lower redshift points was assumed in calculating
the integral luminosity density at $z\sim 1$.
The ground-based point at $z \sim 3$ is in very good
agreement with that from MPD
based solely on the HDF, because the new effective volume is similar to
that assumed in the past, but we find no evidence for a sharp decline
at any redshift larger than $z\sim 1$.
If we were to integrate the luminosity functions over all luminosities instead
of over $L>0.1L^*$, 
the total luminosity density for our $z\sim 3$ and $z\sim 4$ points would be
increased by an additional factor of $1.7^{+0.6}_{-0.2}$, whereas the $z \simlt 1$
points would increase by only $\sim 20$\%.  This is due
to the different faint-end slopes ($\alpha=-1.60$ and $\alpha=-1.3$) that
we have assumed at $z\simgt 3$ and $z\simlt 1$.
It is clear that the real uncertainties in the absolute placement
of any of the points in Figure 9 are much larger than the error bars would
indicate when uncertainty in the shape of the luminosity function
at unobserved magnitudes is taken into account.

The star-formation densities we have calculated so far assume that the
far-UV continua of Lyman-break galaxies have not been extinguished by
dust, but of course the observed colors of these galaxies suggest that this
is unlikely to be the case.  In principle we could correct each galaxy's
magnitude for dust extinction based upon its implied $E(B-V)$, 
construct new luminosity functions at $z\sim 3$ and $z\sim 4$, and then
integrate these to obtain dust-corrected star formation densities.
Unfortunately it is difficult to estimate $E(B-V)$ for galaxies in
our $z\sim 4$ sample, because the ${\cal R}$ and $I$ filters do not
provide a long enough wavelength baseline, and in any case 
the $z\sim 4$ spectroscopic sample is still rather small.  For simplicity
we have instead corrected our $z\sim 3$ and $z\sim 4$ star-formation densities for
extinction by multiplying each by a factor of 4.7, appropriate for
the Calzetti (1997) reddening law with typical $E(B-V) = 0.15$ (cf. Figure 6).  
This correction is reasonably well justified
for the $z\sim 3$ sample, where we know the distribution of $E(B-V)$.  Its
justification at $z\sim 4$ depends upon the assumptions that dust properties are the same 
at $z\sim 3$ and $z\sim 4$ and
that galaxies have the same underlying SEDs; these assumptions are at least
consistent with the limited data, as (for example) Figure 4 shows.
The extinction-corrected luminosity densities are shown in the bottom panel
of Figure 9.  In this figure we have also applied reddening corrections
to the points at other redshifts.  For internal consistency we assumed
the same mean $E(B-V)\simeq 0.15$ and Calzetti extinction law for galaxies
in other samples, so that differences in extinction corrections depend
only on the different rest wavelengths probed at different redshifts.
Under these assumption, the ratio of extinction corrections
at $z>2$ and $z<2$ is set
by the ratio of the effective extinction at $\sim 1500$\AA\ to
that at 2800 \AA, or a factor of approximately 1.7,
implying a multiplicative correction to the $z<2$ points of 2.7.
The implied correction at $z \sim 1$ is consistent with the corrections
deduced by Glazebrook \et (1998) from H$\alpha$ spectroscopy of galaxies from
the Canada-France Redshift Survey sample, as are the corrections implied at $z \sim 0.3$
in comparison to the results of Tresse \& Maddox (1998).  

Because the faint ends of luminosity functions have not been well established
at high redshifts, and the uncertainties in dust extinction are large at all
redshifts, it is probably safest not to read very much into Figure 9.
The point is that there is little compelling evidence, in view of the 
uncertainties involved, for the ``peak'' in 
universal star formation density at $z \sim 1-2$ that has been assumed by many authors.

\section {DISCUSSION}

The precise relationships between the star-formation densities estimated
with various techniques (e.g. UV continua, H$\alpha$, sub-millimeter)
are ambiguous because of the uncertainties in both relative normalization and 
in quantitative interpretation (see, e.g., Kennicutt 1998). 
On the other hand, the consistency of the method we have used to quantify
the UV luminosities being produced by star formation at $z \sim 3$ and $z\sim 4$, coupled with
more secure knowledge of the relevant volumes being probed because of the extensive
spectroscopic sub-samples, allows for a much more robust differential determination  
than has been possible up to this time. While we still regard our result as tentative
because of the relatively small size of the $z\sim 4$ spectroscopic sample, it should
serve as a caution against reaching premature conclusions about the extent to which
the universal star formation history is understood. 

\subsection{Differences Between the Current Sample and the HDF}

There is not strictly speaking any conflict between the results we have presented
and analyses based solely on the Hubble Deep Field (e.g. Madau \et 1996, MPD, etc.).
Our ground-based data are consistent with HDF at both $z\sim 3$
and $z\sim 4$ in the range of apparent magnitudes
that overlap, although this range is very small at $z\sim 4$.  The HDF
provides the only data on fainter LBGs at these redshifts.  In \S 4.3 we re-analyzed
the HDF using improved estimates of the intrinsic spectral-energy distributions
of high-redshift galaxies, and we found, like previous authors, that
the co-moving luminosity density of faint LBGs in the HDF appears to be
roughly 2 times lower at $z\sim 4$ than $z\sim 3$.
This may indicate that the shape of the luminosity function changes
significantly between redshifts $z\sim 4$ and $z\sim 3$---Sawicki \et (1997)
reached this conclusion on somewhat different grounds---but the
good agreement of the the $z\sim 3$ and $z\sim 4$ luminosity functions
at the bright end, where they are the best determined, makes us suspect
instead that the luminosity function may have the same shape at fainter
magnitudes as well, and that the small number of $B_{450}$ dropouts is due
to sample variance in the HDF.  This would not be unexpected given the
strong clustering that LBGs are known to exhibit.

Although the clustering strength of faint LBGs at $z\sim 4$ is not likely
to be measured in the near future, we can explore this hypothesis
further by assuming that the HDF $B_{450}$-dropouts have the same clustering
strength as brighter LBGs at $z\sim 3$.  In our ground-based
sample at $z\sim 3$, the variance of galaxy counts $N$ about the
mean $\mu$ in random regions with angular size similar to the HDF
is larger than the Poisson variance $\langle (N-\mu)^2\rangle = \mu$
by an amount $\mu^2\sigma_{\rm gal}^2$ with $\sigma_{\rm gal}\sim 0.25$.
Because projection effects are comparable in the $z\sim 3$ ground-based
and $z\sim 4$ HDF samples, a similar value of $\sigma_{\rm gal}$
would hold for the HDF $B_{450}$-dropouts if they exhibited the same
clustering as LBGs in our $z\sim 3$ ground-based sample.
If the $\alpha=-1.6$ luminosity function shown in the bottom panel
of Figure 8 were the correct LBG luminosity function at $z\sim 4$,
one would expect an HDF-sized field ($\simeq 5$ arcmin$^{2}$) to contain an average of about 23
$B_{450}$-dropouts with an RMS of 7.5.  The observed number of
$B_{450}$-dropouts, 13, is therefore inconsistent with
this scenario at only the $1.3\sigma$ level.  The formal level of inconsistency
would be smaller still if we had taken proper account of other uncertainties
(e.g. effective volumes, incompleteness corrections, and so on)
in this crude calculation.  The null hypothesis of identical
luminosity function shapes at $z\sim 3$ and $z\sim 4$
cannot be ruled out with much confidence.

%

\subsection{Implications of a Constant UV Luminosity Density at High Redshift}

In calculating the obscuration--corrected luminosity density as a function of redshift
for the LBG samples as in Figure 9, we have implicitly assumed that the mean extinction
corrections for the $z \sim 4$ sample are the same as those at $z \sim 3$. While 
our lever arm for
measuring the UV spectral energy distributions of the $z \sim 4$ galaxies is not
really adequate for independent assessment of their implied ``reddening'', the assumption
that the distribution of galaxy colors is the same at $z \sim 4$ as observed at
$z\sim 3$ has yielded a predicted redshift distribution that is in good agreement with the
information on $N(z)$ at $z \sim 4$. Measurements at longer wavelengths
(e.g., J or possibly even z band) of the $z > 4$ galaxies and a larger spectroscopic sample 
will make this kind of assumption more secure. We have also shown that the typical inferred
obscuration of $z \sim 3$ galaxies is entirely consistent with obscuration estimates based
on independent measurements at smaller redshifts; for these reasons  
we regard it as a reasonable assumption that
the mean obscuration at $z \sim 4$ is not radically different from that at $z \sim 3$. 
A full discussion of the effects of obscuration on the normalization and shape of
the far--UV luminosity function is deferred to another paper (Adelberger \et 1999). 

If the obscuration--corrected far--UV luminosity density and luminosity functions are unchanged between
$\langle z \rangle = 4.13$ and $\langle z \rangle = 3.04$, there are some interesting
implications. 
This would be the first strong evidence 
for a divergence of the universal behavior of star formation as compared to
that of the space density of luminous AGN. It is a general result 
for both
optically selected (Schmidt \et 1995, Kennefick \et 1995, Warren, Hewett, \& Osmer 1994) 
and radio-selected
QSOs (Hook, Shaver, \& McMahon 1998; Shaver \et 1998) that the space density of bright
QSOs decreases by a factor of $\sim 3-4$ between $z \sim3$ and
$z \sim 4$. A similar, or slightly larger, decline is observed in the (radio) luminosity
density of the radio galaxy population in the same range of redshifts (Dunlop 1997). Many have
remarked in retrospect that it would not be surprising if star formation density
and AGN number density followed similar histories, since the formation of
galaxies capable of prodigious star formation and those able to develop massive
central black holes are likely to be correlated to some extent (e.g., Magorrian \et 1998). 
On the other hand, in light of
the data presented here, it is probably not difficult to come up with reasons
why they might not behave in precisely the same way, particularly
at the highest redshifts (c.f. Haehnelt and Rees 1993).  

Very recently, there have been a number of exciting results of galaxy surveys conducted
at wavelengths of 850 $\mu$m using the the SCUBA bolometer array on
the James Clerk Maxwell telescope (e.g., Smail \et 1997, Blain \et 1998, Hughes \et 1998,
Barger \et 1998, Lilly \et 1998, Eales \et 1998), inspiring a great deal of speculation on the implications for
the star formation history of the universe, particularly at high redshift. While it
is not yet clear how much overlap there will be between luminous
sub-millimeter sources and Lyman-break galaxies,
if there is any correlation with the bright end of the (observed) UV-selected galaxy luminosity
function and potential sub-millimeter sources, then it might be expected that
the luminosity density of sub-mm (rest-frame far IR) sources will remain high at $z \sim 4$. On the
other hand, if the brightest sub-millimeter sources are related primarily to
AGN activity (or, similarly, to the same physical processes that result in AGN
activity)-- the only published, positive identification 
of a high redshift sub-mm source indeed has an AGN component (Ivison \et 1998)-- then
one might instead expect a peak in the $z\sim 2-3$ range with a substantial
decline at higher redshift. Ultimately, differences in the redshift distributions
of luminous UV and luminous far--IR sources could prove interesting; however,
a meaningful comparison between the SCUBA results and those presented here is not possible
at present because, as we have shown, the luminosity densities require an accurate
knowledge of the effective volume of a survey and also the luminosity distribution. Neither
is available for the SCUBA sources pending larger spectroscopic samples, no matter how elaborate
a model one constructs. 

\subsection{Comparison With Expectations of Models of Galaxy Formation}

In a comprehensive investigation of the universal star formation history and 
the implications and constraints provided by cosmic backgrounds at various
wavelengths, MPD (see also Calzetti \& Heckman 1998) considered a number of
different star formation histories, including one designed to mimic the 
traditional ``monolithic collapse''. In this model, there is no decline in the luminosity density produced
by star formation at $z >1$, so that the star formation history looks similar to the extinction-corrected
points in Figure 9. In order to reproduce the observed data for the UV luminosity
density in the HDF, it was necessary to add dust extinction whose importance was an
increasing function of redshift. Given the new data, as shown above, it is
possible to accommodate a similar star formation history without requiring that 
the role played by dust change in any way as a function of redshift. As pointed
out by MPD, because of the limited time involved at the highest redshifts,
the only significant impact of having a larger amount of star formation at
high redshift is to exceed the mean metallicity as expressed in the chemical
abundances of high redshift damped Lyman $\alpha$ systems near $z \sim 3$. 
However, it may well be that the DLAs do not, and in fact are not expected to,
trace the same objects as appear to be producing the bulk of the 
high redshift stars and metals (Pettini \et 1999, Somerville, Primack, \& Faber 1998,
Mo, Mao, \& White 1998). Once this constraint is removed, it appears that
more high-redshift star formation is not in strong contradiction with any
of the observable integrated backgrounds (cf. Calzetti \& Heckman 1998).

An interesting question is whether the observed constant star-formation
density between redshifts 3 and 4 is consistent with hierarchical models
for structure formation.  An upper limit to the star-formation density
at a given redshift is the rate at which gas can cool within the virialized
objects at that redshift.  If the Lyman-break galaxies which we observe
are associated with only the most massive virialized dark halos, as we have
argued in Adelberger \et (1998), then bremsstrahlung emission
will likely be a major component of the cooling rate
and supernova feedback will be nearly
negligible.  In this case, assuming self-similar gas distributions,
the hot gas in an observed galaxy would lose energy at a rate
$L_{\rm brems} \propto r_{\rm vir}^3 \rho^2 T^{1/2}$.  The
hierarchical scaling relations $r_{\rm vir}^3 \propto M/(1+z)^3$,
$\rho\propto (1+z)^3$, $T^{1/2}\propto v_c \propto M^{1/3}(1+z)^{1/2}$
(e.g. Kaiser 1986, Navarro, Frenk \& White 1997) then imply 
$L_{\rm brems} \propto M^{4/3} (1+z)^{7/2}$.  Approximating the
rate at which cold gas is produced, $\dot M_{\rm cool}$, as
$L_{\rm brems} \propto \dot M_{\rm cool} v_c^2$ gives
$\dot M_{\rm cool} \propto M^{2/3}(1+z)^{5/2}$.  We can
estimate the ratio of typical galaxy masses in our $z=3$ and $z=4$
samples using the Press-Schechter (1974) model as described (for example)
in Adelberger \et (1998).
For the $\Gamma\sim 0.2$ power spectrum shape suggested by local
measurements, the galaxies in our $z\sim 4$ sample would have typical masses
a factor of 3, 1.3, 2 times lower than galaxies of comparable abundance
in our $z\sim 3$ sample for $\Omega_M=1$, $\Omega_M=0.2$ open,
and $\Omega_M=0.3$ flat.  We would therefore expect our 
samples of galaxies at $z\sim 3$ and $z\sim 4$ to receive newly cooled gas
at comparable rates; using the
expression above, we estimate that $\dot M_{\rm cool}$ is about
1.2, 0.9, 0.7 times higher in $z\sim 3$ sample than our $z\sim 4$
sample---though it would be a mistake to take these numbers too seriously, as our
estimate of $\dot M_{\rm cool}$ is rather crude.  Perhaps a better estimate
of the cooling rate is $\dot M_{\rm cool} \propto M^{5/6}(1+z)^2$, 
proposed (for $\Omega_M=1$) by White \& Frenk (1991), but this
would not substantially alter our conclusion that comparable amounts
of newly cooled gas are available in our $z\sim 3$ and $z\sim 4$ samples.
The point is that two general characteristics of hierarchical models---objects
which form at higher redshift are less massive and more dense than objects
which form at lower redshift---have opposite effects on the rate at which
gas cools in virialized objects, and, over the redshift interval
$z\sim 4$ to $z\sim 3$, these opposite effects largely cancel out
under a range of plausible assumptions.  This suggests that the observed
constant star-formation density between $z\sim 4$ and $z\sim 3$ is by no
means unexpected in hierarchical models---and indeed it was predicted
as early as 1991, long before the relevant observations were made,
by White \& Frenk (1991, Fig. 5).

A separate question is whether the constant star-formation density
is consistent with the prescriptions of semi-analytic models for
turning cooled gas into stars.  The arguments of the previous
paragraph suggest that typical star formation rates in massive halos
will need to scale moderately with halo dark mass $M$ and more strongly
with redshift $z$ to successfully reproduce the constant star formation
density we have observed.  Our uncertainty in the evolution of
the star formation density from $z\sim 4$ to $z\sim 3$ is still
rather large, and we will not press this argument further here;
but in principle it will be possible with observations
such as these to empirically constrain the relationship between
halo dark mass and star formation rate.  This largely
unconstrained relationship has proven in the past to be a fertile source of
free parameters for semi-analytic models; that may not be the case
much longer.

The very steep faint end slope ($\alpha=-1.60$) of the current composite
$z \sim 3$ LBG luminosity function (Figure 8), if taken at face value, suggests
that a large fraction of the luminosity density at $z \sim 3$ could be produced
by objects that are beyond the detection limits for color-selected high redshift
galaxies even in Hubble Deep Field data. However,
in the context of hierarchical galaxy formation models,
if the star formation rates are related to the dark matter halo masses
in a monotonic way (cf. Adelberger \et 1998), then at some point the
luminosity function is expected to become much less steep than the
mass function because of feed-back effects (e.g., Cole \et 1994). For this
reason, it would probably be incorrect to assume that the galaxy far--UV
luminosity function could remain as steep as observed to arbitrarily faint magnitudes. 
The implications for galaxy formation models of the far--UV luminosity function  
and its evolution at high redshift 
will be explored
in much more detail in Adelberger \et 1999. 
Finally, the faint end slope we derive for the $z \sim 3$ LBG luminosity function 
is very similar to that determined 
for far--UV selected galaxies at $z \sim 0.15$, $\alpha=-1.62$ (Treyer \et 1998). 

The current uncertainties in the observed UV luminosity function at $z \simgt 4$ 
(and therefore the uncertainties in epoch-to-epoch comparisons)
will certainly be reduced as more redshifts are obtained, and it is
clearly interesting in light of the discussion above to probe fainter into the $z \sim 4$ luminosity
function with wide-angle surveys. More spectroscopy will allow more
precise estimates of the effective volume imposed by a given set of color
selection criteria, and will allow a better determination of the range of
galaxy properties observed at a given redshift. There is of course the possibility
of extending this type of survey to higher redshifts, as shown by recent
successes in identifying galaxies at $z > 5$ (Dey \et 1998; Weymann \et 1998; Lanzetta,
Yahil, \& Fernandez-Soto 1996). 
The difficulty will come in gathering enough statistics to quantify 
the galaxy population at extreme redshifts, as one will be
forced to still fainter apparent magnitudes (in order to observe the
same range of intrinsic luminosities)  and longer wavelengths for
the follow-up spectroscopy, and the surface densities will begin to be so small
that one will not be able to make optimum use of multi-object spectrographs
on large telescopes. A concern is that galaxies without strong Lyman
$\alpha$ emission at much higher redshifts are likely to be extraordinarily
difficult to confirm spectroscopically. On the other hand, systematic searches
for much higher redshift galaxies, and follow-up spectroscopy, would be
easily within reach with IR cameras and spectrographs on NGST. 

\section{SUMMARY} 

We have presented new photometric data covering 828 square arc minutes, and spectroscopic 
redshifts for 48 of the 244 candidates to $I_{AB}=25.0$, for Lyman-break galaxies in the redshift
range $3.8 \simlt z \simlt 4.5$. These objects were selected using criteria that
facilitate a direct comparison to the properties of the
Lyman-break galaxies in a larger sample covering the range
$2.7 \simlt z \simlt 3.4$.  Besides demonstrating the feasibility of assembling
large samples of $z \simgt 4$ galaxies with ground-based imaging and spectroscopy, 
we have reached the following principal conclusions:

1. The spectroscopic properties of LBGs at $z \simgt 4$ are very similar to those of
the brighter LBGs at $z\sim 3$.  There is a wide range of spectroscopic properties,
and the Lyman $\alpha$ line appears only in absorption for $\sim 50\%$ of the spectroscopically
confirmed objects. We have used the results of the spectroscopy to fine-tune the $z\sim 4$ color
selection criteria to minimize contamination by lower redshift interlopers, and 
to maximize the completeness of the photometrically selected samples. 

2. We see no evidence in our ground-based samples for a significant change
in either the shape or normalization of the LBG luminosity function
between redshifts $\langle z \rangle = 4.13$ and $\langle z \rangle = 3.04$.
Of our results, this is the one which depends upon the fewest assumptions.
It holds for each of the cosmologies we consider ($\Omega=1$, $\Omega=0.2$ open,
$\Omega=0.3$ flat), and is largely independent of our newly estimated incompleteness
corrections.  If there is any large change in the UV-luminosity density
between $z\sim 4$ and $z\sim 3$, it must be due to objects too faint
to be included in our ground-based sample.  We agree with several previous authors
that the HDF provides some evidence that these intrinsically faint LBGs are
rarer at $z\sim 4$ than $z\sim 3$, but the evidence is meager---there are only
13 $B_{450}$ dropouts in the HDF. We have shown that the expected sample variance
for a volume the size of the HDF's renders the apparent deficit of $ z\sim 4$
galaxies statistically insignificant. 

3.  We have developed a working model of the distribution of LBG colors which, when
folded in with our color selection criteria and the observational uncertainties,
successfully reproduces the observed colors as a function of redshift and
the observed redshift distributions in the spectroscopic samples. 
Within the context of this model, the observed far-UV colors imply that a typical
LBG suffers about 1.6 mag. of extinction at 1500\AA\ in the rest frame (cf. Pettini \et 1998), with
a range from zero extinction to $\sim 5$ magnitudes. If we apply an internally
consistent extinction correction to the observed luminosity density points, and
assume that dust plays an equal role at all redshifts,
we find that the emissivity per unit co-moving volume due to star formation
remains {\it essentially flat} for all redshift $z >1$;
hence, it is not clear that the beginning of the epoch of star formation has
yet been identified. 

4. We have presented a new composite $z\sim 3$ far--UV luminosity function,
based on a combination
of data from our ground based survey and from a re-analysis of the HDF data
based on our improved estimate of the survey effective volumes.  
The luminosity function at $z\sim 3$ is well fit by a Schechter function
with a faint end
slope of $\alpha = -1.60\pm 0.13$, considerably steeper than previous
estimates.  A similar fit results from using the ground-based sample alone. 
Such a steep faint end slope allows for the possibility that a
large fraction of the luminosity density at high redshift lies beyond even
the detection limits in the Hubble Deep Field. The lack of adequate
constraints on the behavior
of the LBG luminosity functions at faint magnitudes, particularly at
$z \simgt 4$, is an additional, possibly large, source of uncertainty in
estimating the total luminosity density at high redshifts.  

5. The similarity of the (observed) $z \sim 4$ and $z \sim 3$ bright-end luminosity
functions indicates a significant difference in the evolutionary behavior
of universal star formation as compared to the space density of luminous AGNs. 

6. The apparent constancy of the UV luminosity function between $z \sim 3$ and
$z \sim 4$ has interesting implications for the redshift distribution
for the newly discovered population of luminous sub-millimeter sources, 
and for the star formation prescriptions in semi-analytic models of galaxy formation.

\bigskip
\bigskip

We are grateful to the many people responsible for
building and maintaining the W.~M.~Keck telescopes and the Low Resolution
Imaging Spectrograph. We also thank the dedicated staffs of the Palomar, 
Keck, Cerro Tololo, and La Palma observatories for making these
difficult observations possible.   
Software by J. Cohen, A. Phillips, and P. Shopbell helped
in slit-mask design and alignment. We are also grateful to the
referee, M. Shull, whose constructive comments improved the presentation of the results. 
CCS acknowledges support from the U.S. National Science Foundation through
grant AST 94-57446, and from the David and Lucile Packard Foundation. 
MG has been supported through grant HF-01071.01-94A from the Space Telescope
Science Institute, which is operated by the Association of Universities for
Research in Astronomy, Inc. under NASA contract NAS 5-26555.

\bigskip
\newpage
\newpage
\begin{deluxetable}{lcrcrc}
\tablewidth{0pc}
\scriptsize
\tablecaption{$z \simgt 4$ Lyman-Break Galaxy Fields}
\tablehead{
\colhead{Field Name} & \colhead{Field Center}  & \colhead{Area} & \colhead{\# Candidates\tablenotemark{a}} & 
\colhead{$z_{\rm true}$\tablenotemark{b}} & \colhead{$z_{\rm false}$\tablenotemark{c}} \nl
\colhead{} & \colhead{(J2000)} & \colhead{(arc min)$^{2}$} & \colhead{} & \colhead{} & \colhead{}  
 } 
\startdata
CDFa       & 00 53 23.7  +12 34 00 & 78 & 22/21 & 6/5 & 0/0 \nl 
CDFb       & 00 53 43.0  +12 25 15 & 82 & 26/22 & 8/8 & 1/0 \nl
B2 0902+34 & 09 05 30.2  +34 07 55  & 40 & 13/8 & 3/3 & 4/1 \nl
HDF & 12 36 52.3  +62 12 59 & 75 & 26/20 & 2/2 & 1/0 \nl
3C324 & 15 49 49.6  +21 29 07 & 42 & 15/10 & 2/2 & 4/2 \nl
DSF1550 & 15 51 02.4  +07 22 34 & 180 & 61/58 & 7/7 & 1/1 \nl
SSA22a & 22 17 34.2  +00 15 01 & 78 & 28/24 & 5/5 & 9/4 \nl
SSA22b & 22 17 34.2  +00 06 18 & 78 & 13/11 & 3/3 & 1/0 \nl
DSF2237a & 22 40 08.5  +11 52 34 & 83 & 25/21 & 6/6 & 1/1 \nl
DSF2237b & 22 39 34.3  +11 51 44 & 82 & 16/12 & 6/5 & 2/1 \nl\nl
TOTAL & & 828 & 244/207 & 48/46 & 25/12\nl
\enddata
\tablenotetext{a}{The first entry applies to the initial, broad color selection
window; the second entry is the number of candidates remaining after
applying the ${\cal R}-I \le 0.6$ color criterion and the $I\ge 23.0$ apparent
magnitude cut (see text). }
\tablenotetext{b}{Number of spectroscopic redshifts of $z \sim 4$ galaxies,
before and after application of the additional color and magnitude criteria.} 
\tablenotetext{c}{Number of spectroscopic redshifts of interloper galaxies
($z << 4$, before and after application of additional color and magnitude
criteria).}
\end{deluxetable}
\newpage
\begin{deluxetable}{lccccccl}
\tablewidth{0pc}
\scriptsize
\tablecaption{Lyman-Break Galaxy Candidate Spectroscopic Results}
\tablehead{
\colhead{Object Name} & \colhead{RA (J2000)} & \colhead{Dec (J2000)} &\colhead{$I_{\rm AB}$} & \colhead{$({\cal R}-I)_{\rm AB}$} & \colhead{$(G-{\cal R})_{\rm AB}$} & 
 \colhead{$z$} 
 } 
\startdata
CDFa-G1 & 00 53 33.25 & +12 32 07.3 & 23.59 & 0.89 & $>$3.72 & 4.815 \nl
CDFa-GD3 & 00 53 26.93 & +12 30 45.6 & 24.92 & -0.18 & 2.47 & 4.050 \nl
CDFa-GD4 & 00 53 36.20 & +12 31 09.1 & 23.49 & 0.28 & 2.07 & 4.189 \nl
CDFa-GD7 & 00 53 35.60 & +12 31 44.2 & 23.55 & 0.50 & 3.71 & 4.605 \nl
CDFa-GD9 & 00 53 36.13 & +12 32 50.2 & 24.80 & 0.18 & 2.45 & 3.864 \nl
CDFa-GD14 & 00 53 24.66 & +12 35 31.3 & 24.74 & 0.09 & 2.02 & 3.613 \nl
CDFb-G1 & 00 53 35.69 & +12 21 06.2 & 24.85 & -0.17 & $>$3.24 & 4.077 \nl
CDFb-G5 & 00 53 51.26 & +12 24 21.3 & 24.84 & 0.38 & $>$2.77 & 4.486 \nl
CDFb-GD3 & 00 53 45.68 & +12 21 27.3 & 24.39 & 0.11 & 2.06 & 3.694 \nl
CDFb-GD9 & 00 53 44.41 & +12 21 56.8 & 23.80 & 0.31 & 2.17 & 3.777 \nl
CDFb-GD10 & 00 53 50.58 & +12 23 39.7 & 24.20 & 0.25 & 2.75 & 4.070 \nl
CDFb-GD12 & 00 53 40.69 & +12 24 29.3 & 24.92 & 0.24 & 2.32 & 3.469 \nl
CDFb-GD13 & 00 53 46.59 & +12 24 38.4 & 24.71 & 0.07 & 2.11 & 3.761 \nl
CDFb-GD14 & 00 53 36.49 & +12 24 34.2 & 24.50 & 0.27 & 2.35 & 4.253 \nl
B20902-G1 & 09 05 22.10 & +34 07 51.0 & 24.64 & -0.01 & $>$3.13 &  4.318 \nl
B20902-GD5 & 09 05 18.93 &  +34 07 51.8 &24.67 & -0.31 & 2.68 &  4.039 \nl
B20902-GD11 & 09 05 34.54 &  +34 10 26.0 & 24.23 & -0.23 & 3.44 &  4.204 \nl
HDF-G4 & 12 37 20.56 &  +62 11 06.5 &24.32 & -0.04 & $>$3.70 & 4.421 \nl
HDF-GD4 & 12 37 10.65 &  +62 09 35.8 &24.99 & 0.06 & 2.76 &  4.129 \nl
3C324-G2 & 15 49 48.65  & +21 26 49.8 & 24.61 & 0.39 & $>$2.63 & 4.434 \nl
3C324-GD2 & 15 49 41.12  & +21 27 34.5 &24.63 & 0.45 & 2.58 &  4.071 \nl
DSF1550-G4 & 15 50 36.23  & +07 16 54.6 &24.54 & 0.44 & $>$3.13 & 4.133 \nl
DSF1550-GD10 & 15 50 56.77  & +07 15 42.1 &24.83 & -0.47 & 2.25  & 3.914 \nl
DSF1550-GD18 & 15 50 39.41 &  +07 16 44.5 &24.75 & -0.27 & 2.22  & 4.144 \nl
DSF1550-GD22 & 15 50 37.14 &  +07 17 52.8 &24.40 & -0.17 & 2.10 &  3.894 \nl
DSF1550-GD23 & 15 50 44.13 &  +07 18 20.1 &23.97 & 0.15 & 2.33 &  4.053 \nl
DSF1550-GD30 & 15 50 48.31 &  +07 20 45.5 &24.50 & 0.16 & 2.30 &  4.299 \nl
DSF1550-GD32 & 15 50 45.48 &  +07 20 54.3 &24.46 & -0.21 & 2.34 &  4.176 \nl
SSA22a-G3 & 22 17 30.79 & +00 12 50.9 & 24.65 & 0.38 & $>$2.83 & 4.527 \nl
SSA22a-G6 & 22 17 48.39 & +00 14 34.5 & 24.72 & -0.06 & $>$3.02 & 3.710 \nl
SSA22a-G11 & 22 17 39.43 & +00 17 46.0 & 24.91 & 0.58 & 2.80 & 4.390 \nl
SSA22a-GD11 & 22 17 42.78 & +00 16 18.1 & 24.59 & 0.11 & 2.23 & 4.397 \nl
SSA22a-GD12 & 22 17 48.39 & +00 16 25.9 & 24.24 & 0.17 & 2.18 & 4.114 \nl
SSA22b-GD7 & 22 17 32.40 &  +00 05 32.5 &24.47 & 0.56 & 2.95 &  3.895 \nl
SSA22b-GD9 & 22 17 42.49 &  +00 07 26.4 &23.81 & 0.16 & 2.14 &  3.856 \nl
SSA22b-GD10 & 22 17 33.80 &  +00 08 48.3 &23.36 & 0.40 & 2.70 &  4.086 \nl
DSF2237a-G3 & 22 40 03.64 & +11 49 12.6 & 24.75 & -0.08 & $>$3.13 & 4.400 \nl
DSF2237a-G7 & 22 40 13.09 & +11 53 32.2 & 24.70 & 0.13 & $>$3.28 & 3.838 \nl
DSF2237a-G10 & 22 40 18.80 & +11 55 44.4 & 24.44 & 0.52 & $>$3.27 & 4.467 \nl
DSF2237a-GD3 & 22 40 06.88 & +11 49 17.8 & 24.70 & 0.14 & 2.11 & 3.964 \nl
DSF2237a-GD7 & 22 40 04.34 & +11 52 34.7 & 24.71 & -0.15 & 2.91 & 4.453 \nl
DSF2237a-GD10 & 22 40 03.87 & +11 53 57.5 & 24.89 & -0.02 & 2.17 & 4.189 \nl
DSF2237b-G2 & 22 39 39.11 &  +11 49 20.5 &24.33 & 0.72 & $>$3.35 & 4.178 \nl
DSF2237b-G4 & 22 39 25.52 &  +11 50 39.2 &24.23 & 0.38 & $>$3.74 & 4.492 \nl
DSF2237b-G12 & 22 39 33.25 &  +11 55 24.9 &24.98 & 0.11 & $>$3.15 & 4.375 \nl
DSF2237b-GD6 & 22 39 33.52 &  +11 51 54.6 &24.98 & -0.06 & 2.03 &  4.486 \nl
DSF2237b-GD11 & 22 39 29.09 &  +11 55 47.0 &24.81 & 0.40 & 2.97 &  3.679 \nl
DSF2237b-GD12 & 22 39 30.35 &  +11 55 54.5 &24.28 & 0.19 & 2.24 &  3.868 \nl
           &                &               &      &      &      &  \nl
CDFb-GD4 & 00 53 30.10 & +12 21 24.1 & 22.72 & 0.98 & 4.01 & 0.000 \nl
B20902-GD2 & 09 05 37.62 &  +34 05 20.3 &23.98 & 0.57 & 2.94 &  1.069 \nl
B20902-GD3 & 09 05 25.20 &  +34 06 11.1 &22.72 & 0.96 & 3.47 &  0.842 \nl
B20902-GD7 & 09 05 29.88 &  +34 09 22.0 &22.42 & 0.89 & 3.59 &  0.651 \nl
B20902-GD8 & 09 05 30.96 &  +34 09 31.7 &23.75 & 0.70 & 3.13 &  0.944 \nl
HDF-GD5 & 12 37 24.71 &  +62 09 46.9 &22.96 & 0.58 & 3.20 &  1.010 \nl
3C324-G1 & 15 49 50.07 &  +21 26 23.1 &24.35 & 0.78 & $>$3.12 & 0.945 \nl
3C324-G6 & 15 49 46.57 &  +21 29 38.0 &24.57 & 0.51 & $>$2.86 & 0.941 \nl
3C324-G8 & 15 49 57.19 &  +21 30 12.8 &24.72 & 0.56 & $>$2.84 & 0.964 \nl
3C324-GD5 & 15 49 56.95 &  +21 31 00.2 &22.85 & 0.97 & 3.54 &  0.809 \nl
DSF1550-GD28 & 15 50 50.36 &  +07 20 16.0 &23.25 & 0.27 & 2.05 &  0.961 \nl
SSA22a-G1 & 22 17 44.68 & +00 12 05.6 & 23.97 & 0.75 & $>$3.09 & 1.061 \nl
SSA22a-G4 & 22 17 23.38 & +00 13 14.4 & 23.85 & 0.72 & $>$3.09 & 1.243 \nl
SSA22a-G8 & 22 17 28.77 & +00 15 49.2 & 24.31 & 0.11 & 3.11 & 1.152 \nl
SSA22a-GD5 & 22 17 42.29 & +00 13 23.6 & 23.76 & 0.58 & 3.06 & 0.630 \nl
SSA22a-GD7 & 22 17 33.89 & +00 14 50.8 & 24.17 & 0.64 & 2.89 & 1.109 \nl
SSA22a-GD8 & 22 17 50.49 & +00 15 18.2 & 23.49 & 0.23 & 2.14 & 0.339 \nl
SSA22a-GD13 & 22 17 41.93 & +00 16 44.0 & 24.18 & 0.24 & 2.19 & 0.670 \nl
SSA22a-GD14 & 22 17 29.16 & +00 17 11.9 & 24.78 & 0.18 & 2.55 & 0.740 \nl
SSA22a-GD15 & 22 17 36.56 & +00 17 46.8 & 24.86 & 0.22 & 2.04 & 1.238 \nl
SSA22a-GD17 & 22 17 37.62 & +00 18 51.3 & 22.02 & 0.90 & 3.33 & 0.785 \nl
SSA22b-G3 & 22 17 42.57 &  +00 02 51.6 &23.44 & 0.82 & $>$3.53 & 1.0 \nl
DSF2237a-GD6 & 22 40 06.05 & +11 50 47.0 & 23.28 & 0.52 & 2.66 & 0.861 \nl
DSF2237b-GD1 & 22 39 32.24  & +11 47 38.0 &24.76 & 0.37 & 2.87 &  1.0 \nl
DSF2237b-GD3 & 22 39 26.10 &  +11 50 19.0 &21.15 & 0.22 & 2.04 &  0.499 \nl
\enddata
\end{deluxetable}
\newpage
\begin{deluxetable}{lcccc}
\tablewidth{0pc}
\scriptsize
\tablecaption{Lyman-Break Galaxy Surface Density and Effective Survey Volumes, 
$z \sim 3$\tablenotemark{a}}
\tablehead{
\colhead{AB Mag Range\tablenotemark{b}} & \colhead{$\Sigma(z=3)$\tablenotemark{c}} & 
\colhead{$V_{\rm eff}(1,0)$\tablenotemark{d}} & 
\colhead{$V_{\rm eff}(0.2,0)$\tablenotemark{d}} & \colhead{$V_{\rm eff}(0.3,0.7)$\tablenotemark{d}} 
}
\startdata
22.5-23.0  &  $0.002\pm0.001$ & 120 & 448 & 471 \nl
23.0-23.5  &  $0.021 \pm 0.004$ & 120 & 448 & 471 \nl 
23.5-24.0  &  $0.087 \pm 0.011$ & 117 & 437 & 459 \nl 
24.0-24.5  &  $0.194 \pm 0.016$ & 112 & 418 & 440 \nl 
24.5-25.0  &  $0.380 \pm 0.023$ & 97  & 362 & 381 \nl
25.0-25.5  &  $0.495 \pm 0.049$ & 67  & 250 & 263 \nl 
\enddata
\tablenotetext{a}{Observed surface density}
\tablenotetext{b}{${\cal R}$ magnitudes}
\tablenotetext{c}{Objects per arc min$^{2}$ in 0.5 magnitude interval for $z \sim 3$ sample.
Each bin has been corrected for contamination by interlopers based on the spectroscopic
sample.
The errors reflect both Poisson counting and field-to-field variations} 
\tablenotetext{d}{Effective survey volume per square arc minute for galaxies
in each range of apparent magnitude, in units of $h^{-3}$ Mpc$^{3}$. 
The numbers in parentheses indicate the assumed cosmology, with ($\Omega_m$, $\Omega_{\Lambda}$).
}
\end{deluxetable}
\newpage
\begin{deluxetable}{lcccc}
\tablewidth{0pc}
\scriptsize
\tablecaption{Lyman-Break Galaxy Surface Density and Effective Survey Volumes, 
$z \sim 4$\tablenotemark{a}}
\tablehead{
\colhead{AB Mag Range\tablenotemark{b}} & \colhead{$\Sigma(z=4)$\tablenotemark{c}} & 
\colhead{$V_{\rm eff}(1,0)$\tablenotemark{d}} & 
\colhead{$V_{\rm eff}(0.2,0)$\tablenotemark{d}} & \colhead{$V_{\rm eff}(0.3,0.7)$\tablenotemark{d}} 
}
\startdata
23.0-23.5  &  $0.004 \pm 0.002$ & 129 & 619 & 537 \nl 
23.5-24.0  &  $0.014 \pm 0.006$ & 129 & 619 & 537 \nl 
24.0-24.5  &  $0.063 \pm 0.013$ & 118 & 566 & 491 \nl 
24.5-25.0  &  $0.131 \pm 0.015$ & 74  & 355 & 308 \nl
\enddata
\tablenotetext{a}{Observed surface density}
\tablenotetext{b}{$I$ magnitudes}
\tablenotetext{c}{Objects per arc min$^{2}$ in 0.5 magnitude interval for $z \sim 4$ sample.
Each bin has been corrected for contamination by interlopers based on the spectroscopic
sample.
The errors reflect both Poisson counting and field-to-field variations} 
\tablenotetext{d}{Effective survey volume per square arc minute, in units of $h^{-3}$ Mpc$^{3}$.
The numbers in parentheses indicate the assumed cosmology, with ($\Omega_m$, $\Omega_{\Lambda}$).}
\end{deluxetable}
\newpage
\begin{deluxetable}{lccc}
\tablewidth{0pc}
\scriptsize
\tablecaption{Observed Ultraviolet Luminosity Densities}
\tablehead{
\colhead{Cosmological Model} &  
 \colhead{${\rm log}\rho_{UV}(z=3)$\tablenotemark{a}} & 
\colhead{${\rm log}\rho_{UV}(z=4)$\tablenotemark{a}} & \colhead{$\rho_{UV}(z=3)/\rho_{UV}(z=4)$} 
}
\startdata
$\Omega_m=1.0$, $\Omega_{\Lambda}=0$ &  $26.05\pm0.07$ & $25.96\pm0.10$ & $1.23\pm 0.38$ \nl  
$\Omega_m=0.2$, $\Omega_{\Lambda}=0$ &  $25.74\pm0.07$ & $25.74\pm0.10$ & $1.02\pm 0.31$ \nl 
$\Omega_m=0.3$, $\Omega_{\Lambda}=0.7$ &  $25.75\pm0.07$ & $25.70\pm0.10$ & $1.13\pm 0.35$ \nl
\enddata
\tablenotetext{a}{UV luminosity density in units of log (ergs s$^{-1}$ $Hz^{-1}$ h Mpc$^{-3}$) .
These numbers are integrated only over  
the range of luminosities in common to the two ground-based samples; no extrapolation to fainter
magnitudes using HDF data or an assumed luminosity function has been included here.}
\end{deluxetable}
\newpage
\begin{figure}
\figurenum{1a}
\plotone{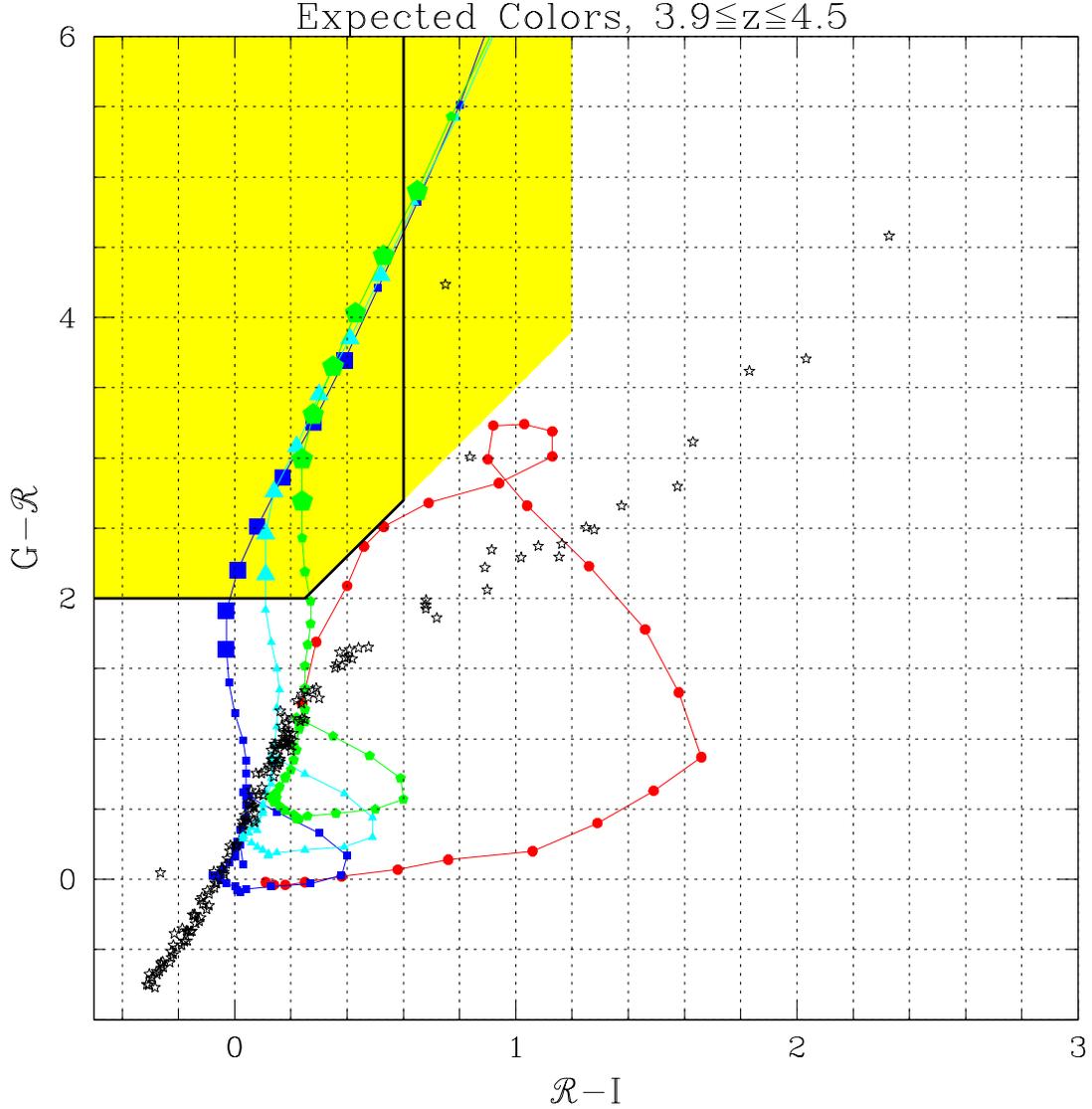}
\caption{Expected locus of colors versus redshift in the adopted 3-band
photometric system, for the $z \sim 4$ sample. The shaded region indicates
that used for our initial spectroscopic sample, while the region enclosed
with the bold lines is that used for the statistical samples (see text for
discussion). 
The positions in the color--color diagrams expected for
3 different degrees of reddening (E(B-V)=0,0.15, and 0.30) are shown
with squares, triangles, and pentagons, respectively. Redshifts in
the range $3.9-4.5$ are indicated with enlarged symbols.
Also shown (solid dots) is the locus for unevolved early type galaxies, ranging from
z=0.1 to z=2.5 in steps of 0.1. Note that early type galaxies encroach on our
selection window for $z \sim 0.5-1$. The ``star'' symbols represent the colors
for Galactic stars.
}
\end{figure}
\begin{figure}
\figurenum{1b}
\plotone{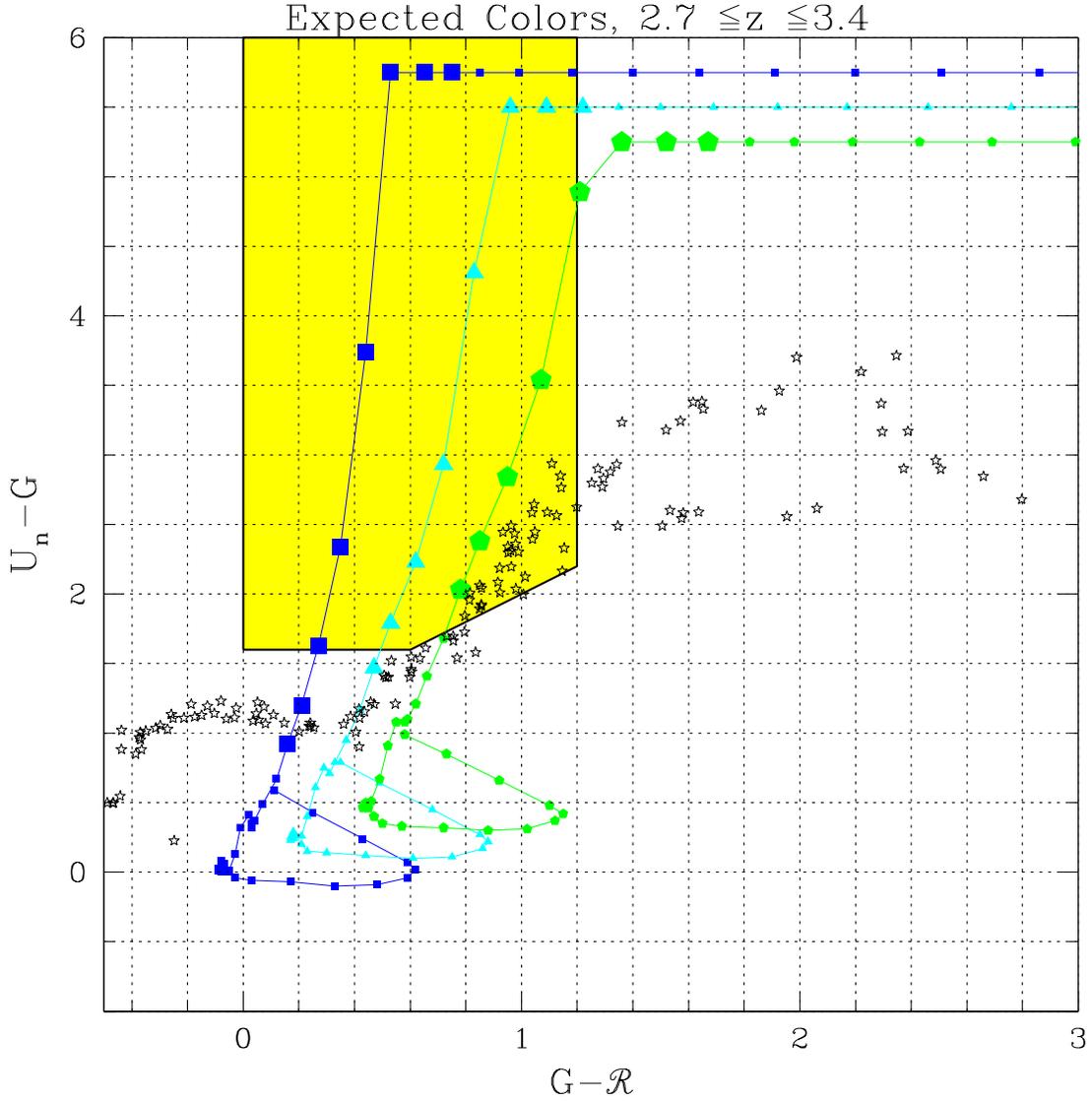}
\caption{Same as a), for the $z \sim 3$ Lyman-break galaxy selection criteria. 
Models for which it is predicted that $U_n -G>6.0$ are shown with the color ``clipped''
at a constant $U_n-G$
in order to display the expected $G-{\cal R}$ colors on the plot. Redshifts from 
2.7 to 3.4 are indicated with the enlarged symbols, where again the squares
correspond to an unreddened model SED, the triangles to one having
$E(B-V)=0.15$ with the Calzetti reddening law, and the pentagons to
one having $E(B-V)=0.30$. Note that for an object having the ``typical''
colors (triangles), the selection window is expected to include objects in the full
range $2.7 \le z \le 3.4$, whereas for the unreddened models one expects only
objects toward the higher redshift end of the range, and for the $E(B-V)=0.3$
tracks, one sees preferentially the lower redshift end of the range. 
}
\end{figure}
\begin{figure}
\figurenum{2}
\plotone{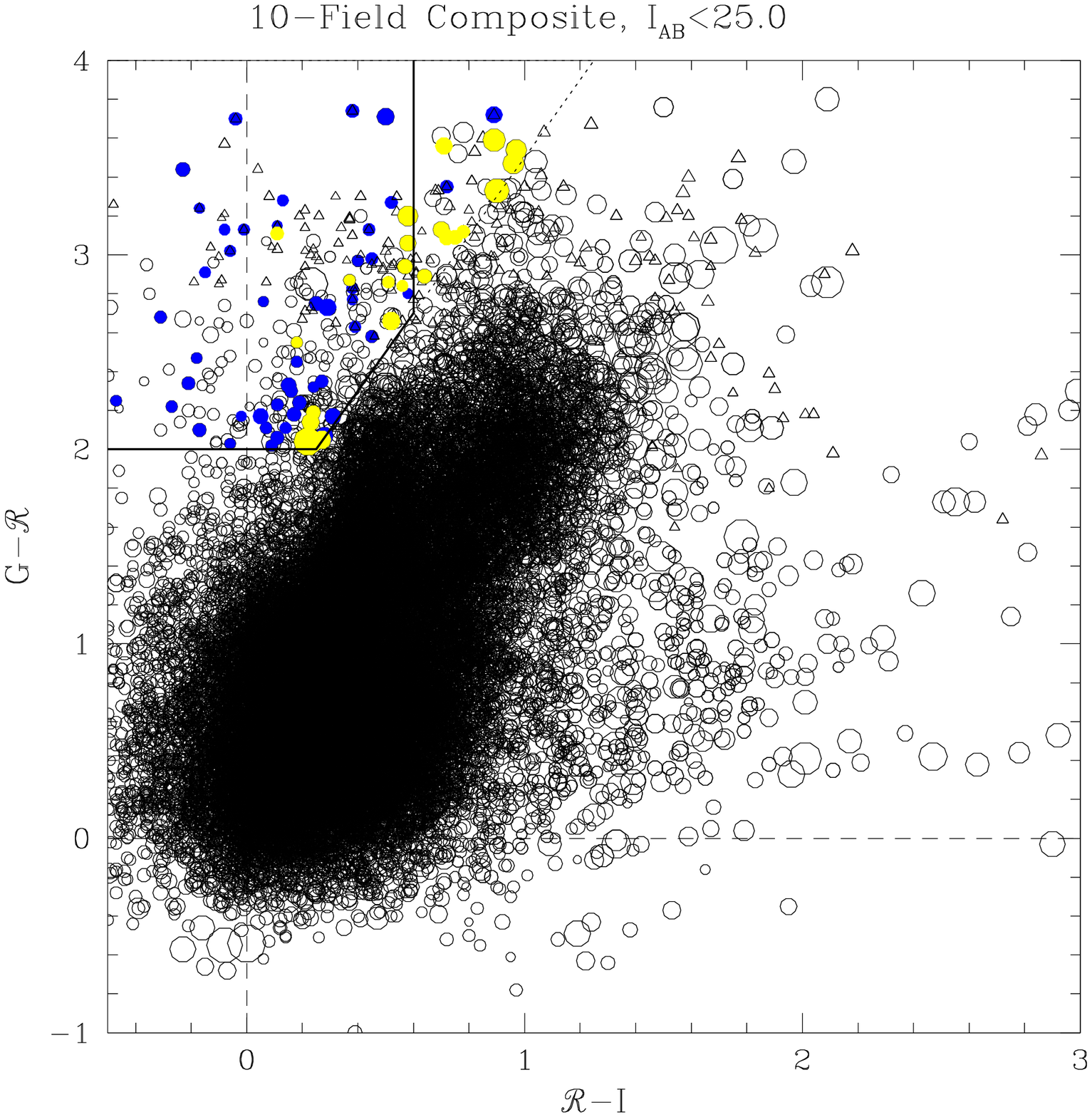}
\caption{10-Field composite color--color diagram, where the spectroscopic G-band
break object selection region is shown enclosed with a dotted line style, and the region used for the statistical
samples is indicated with the heavy line type. There are approximately 29,000 objects 
represented here, of which 207 satisfy the primary color selection criteria. Symbol size scales
inversely with apparent magnitude, and triangles represent objects with only limits
on the $G-{\cal R}$ color. 
Filled symbols show
objects with spectroscopic redshifts, where lighter shading indicates 
``interloper'' galaxies, and darker shading indicates objects with spectroscopic
redshifts in the range $3.7 < z < 4.8$. 
}
\end{figure}
\begin{figure}
\figurenum{3}
\plotone{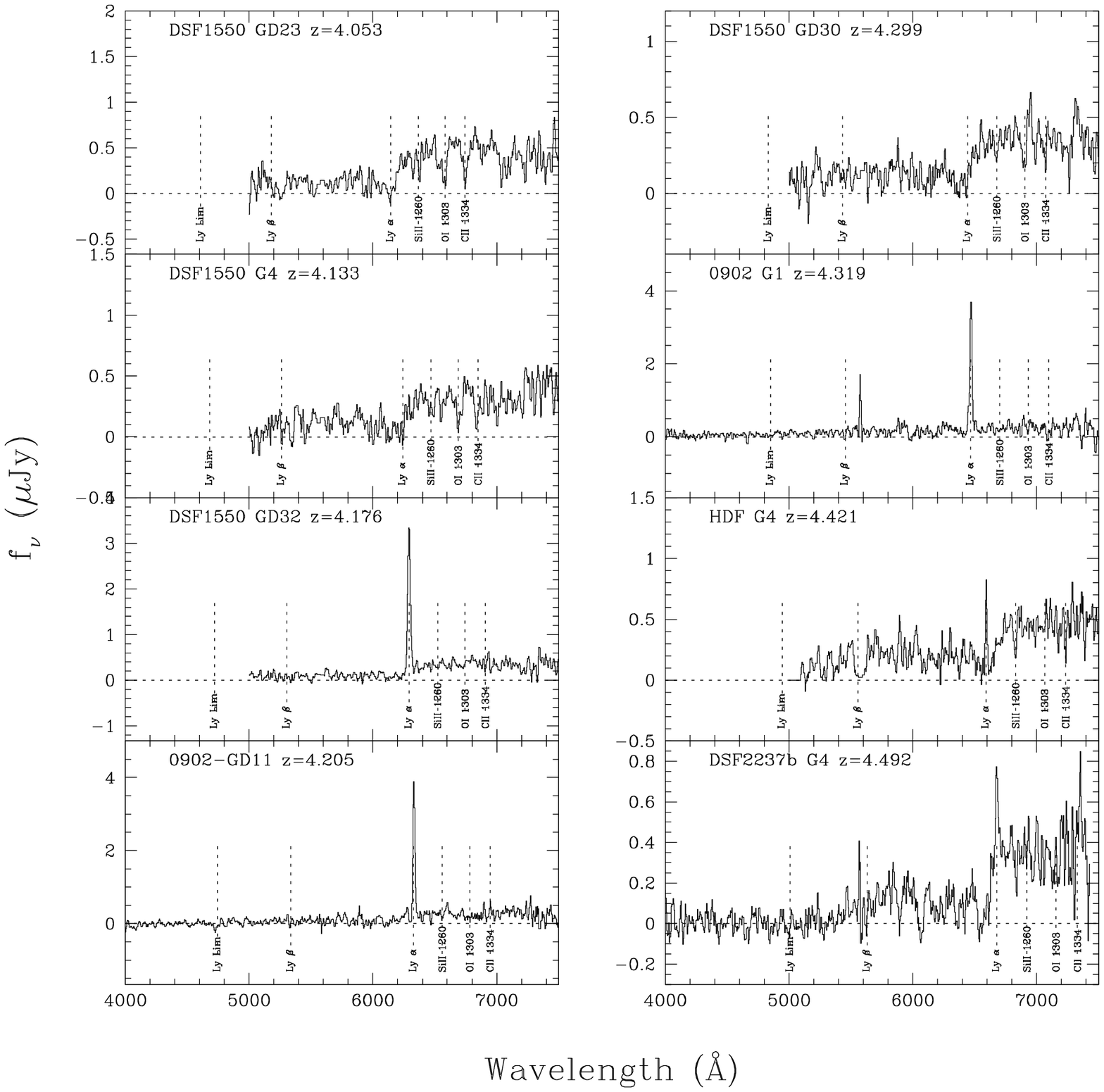}
\caption{Example spectra of G-band break objects. Note that, as for the $z \sim 3$ sample,
the $z\sim 4$ Lyman-break galaxies have a widely varying Lyman $\alpha$ line strength,
from strong emission lines, to very strong and broad absorption. Overall, the spectra
are very similar to the $z \sim 3$ objects at correspondingly bright UV luminosities.} 
\end{figure}
\begin{figure}
\figurenum{4}
\plotone{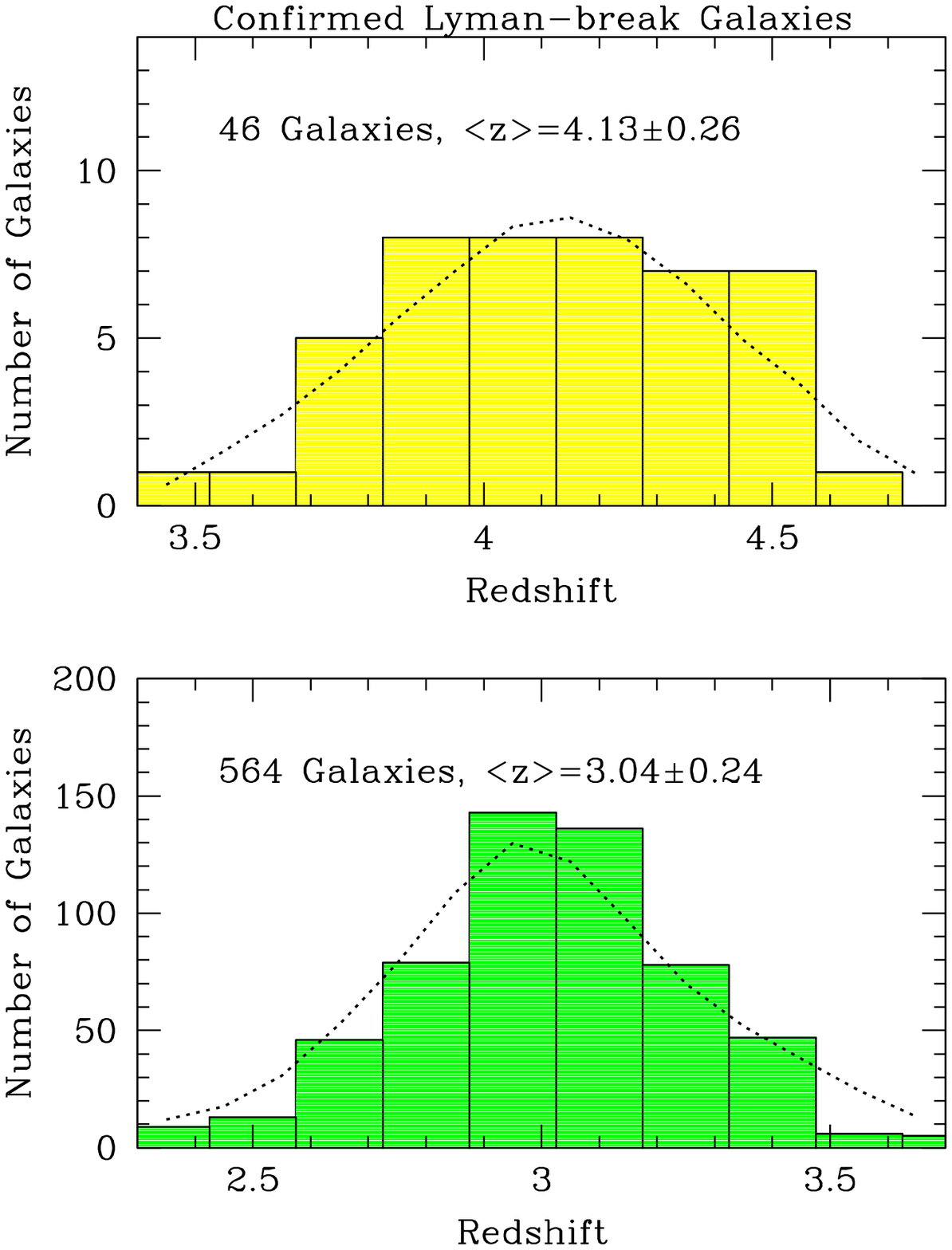}
\caption{ Current histogram of $z\sim 4$ Lyman-break galaxies with spectroscopic
redshifts (upper panel) and $z \sim 3$ LBGs (lower panel). The mean and standard
deviations of the distributions are indicated.
The dotted curves are the predicted redshift distributions for
our working model (see text) that simultaneously reproduces the observed color
distribution as a function of redshift. The curve in the top panel assumes that
the intrinsic color distribution of LBGs is the same in the $z \sim 4$ sample as
measured in the much--larger $z\sim 3$ sample.  
}
\end{figure}
\begin{figure}
\figurenum{5}
\plotone{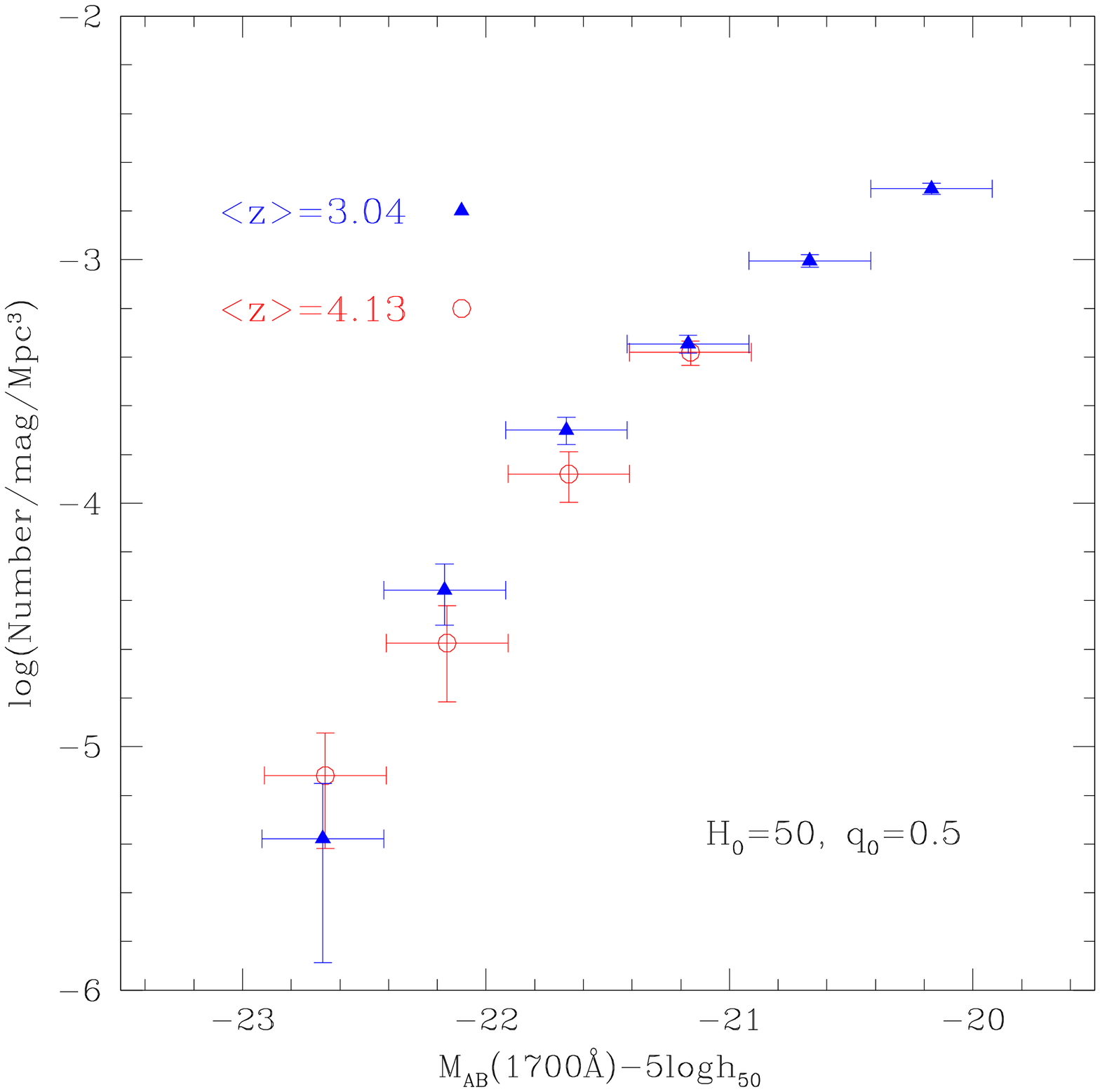}
\caption{A comparison of the luminosity functions
for the two Lyman-break galaxy samples, using the effective sample volumes in Tables 3 and 4, 
and assuming
$\Omega =1$ and $H_0=50$ \kms Mpc$^{-1}$; other cosmologies considered would result
in a shift of the $z \sim 4$ luminosity function to slightly brighter magnitudes relative
to $z \sim 3$ (see Table 5 for the effect on the relative luminosity densities).
}
\end{figure}
\begin{figure}
\figurenum{6}
\plotone{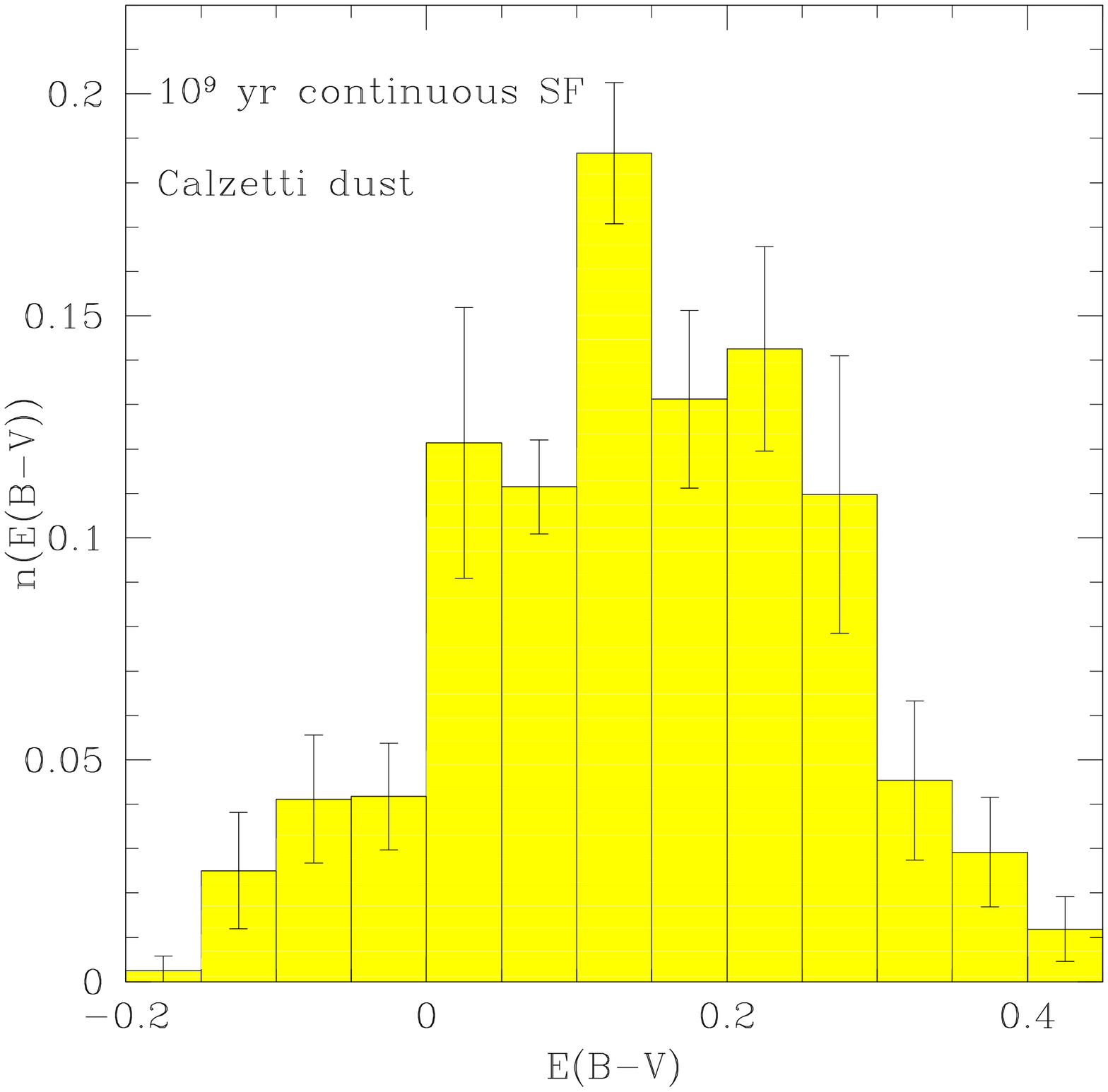}
\caption[f6.eps]{The inferred intrinsic distribution of galaxy colors for LBGs at
$z \sim 3$. The error bars reflect the field-to-field differences in
the color distribution. The E(B-V) values are inferred relative to a spectral
template that assumes constant star formation for $\sim 10^9$ years, internal
Lyman continuum opacity, and opacity due to the intergalactic medium, as discussed
in \S 3. }
\end{figure}
\begin{figure}
\figurenum{7}
\plottwo{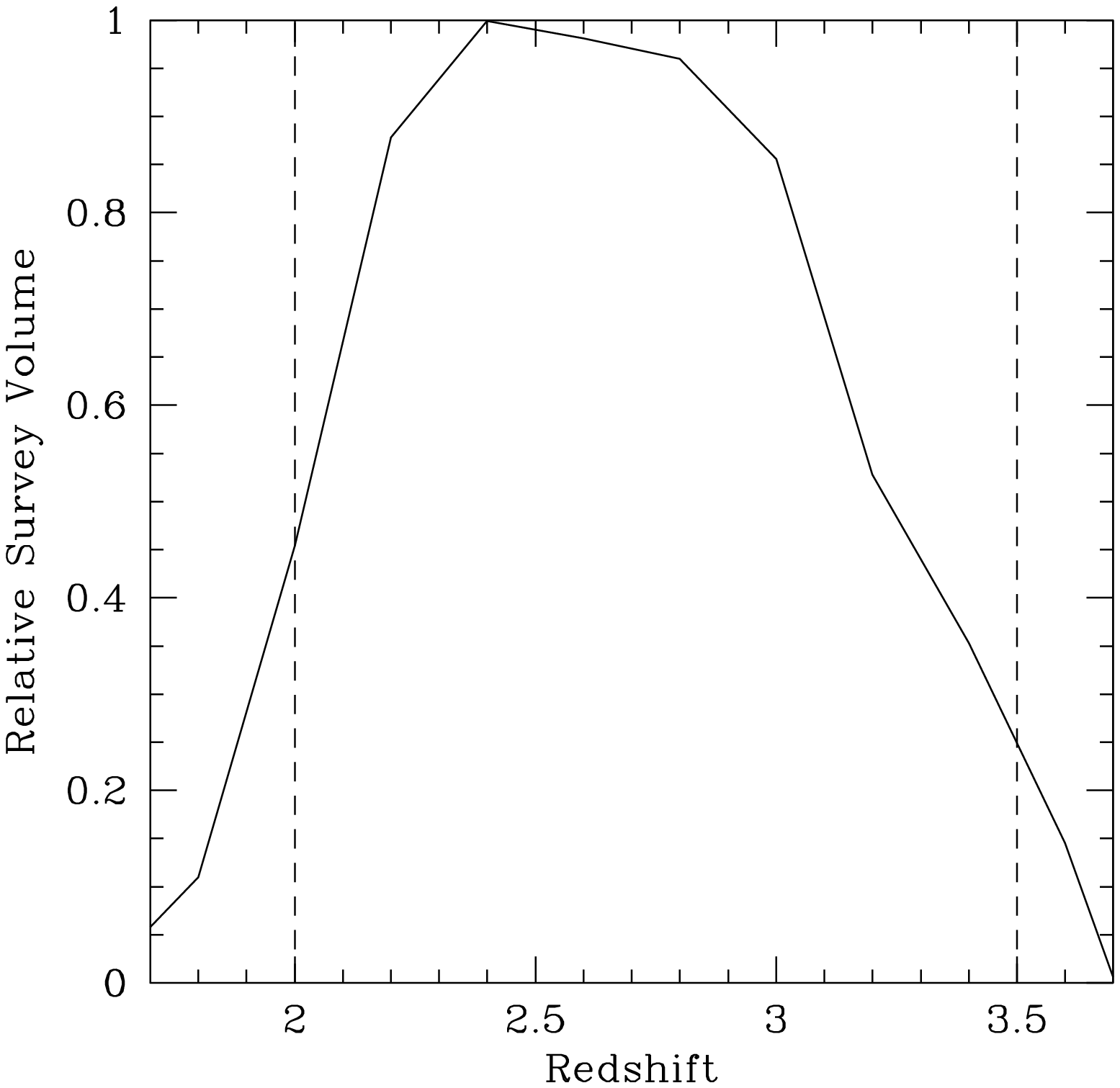}{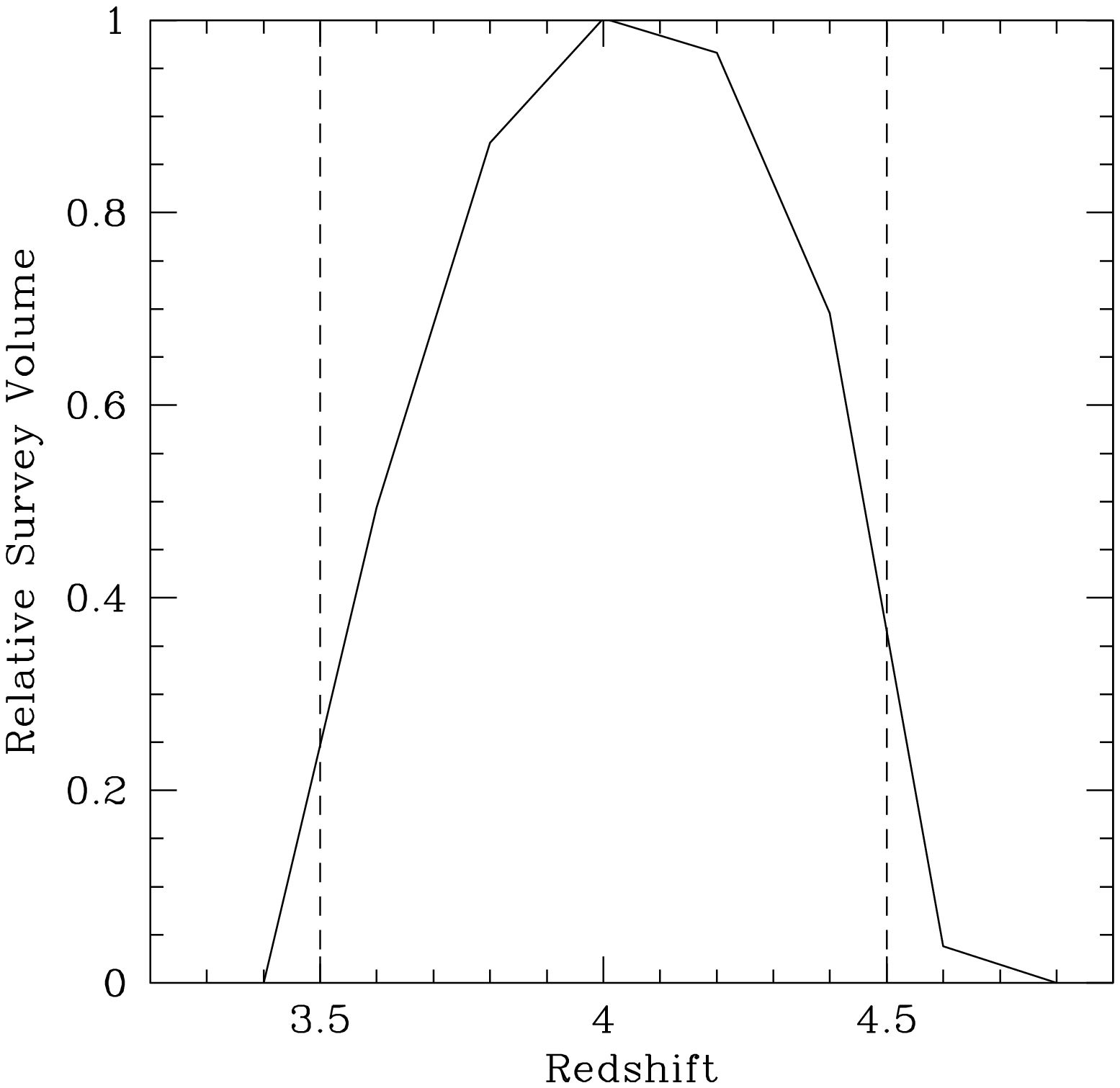}
\caption{Plots showing the expected effective redshift distribution for LBG samples
chosen using the HDF filter system and the selection criteria given in
the text 
(cf. Madau \et 1996), if the true density of galaxies remained constant over the
redshift interval and if galaxies had the distribution of colors that is found
in the ground-based spectroscopic sample. The left panel is for F300W ``dropouts'', and
the right panel is for F450W ``dropouts''. The dashed lines indicate the
redshift selection functions assumed by Madau \et (1996, 1998). }
\end{figure}  
\begin{figure}
\figurenum{8}
\plotone{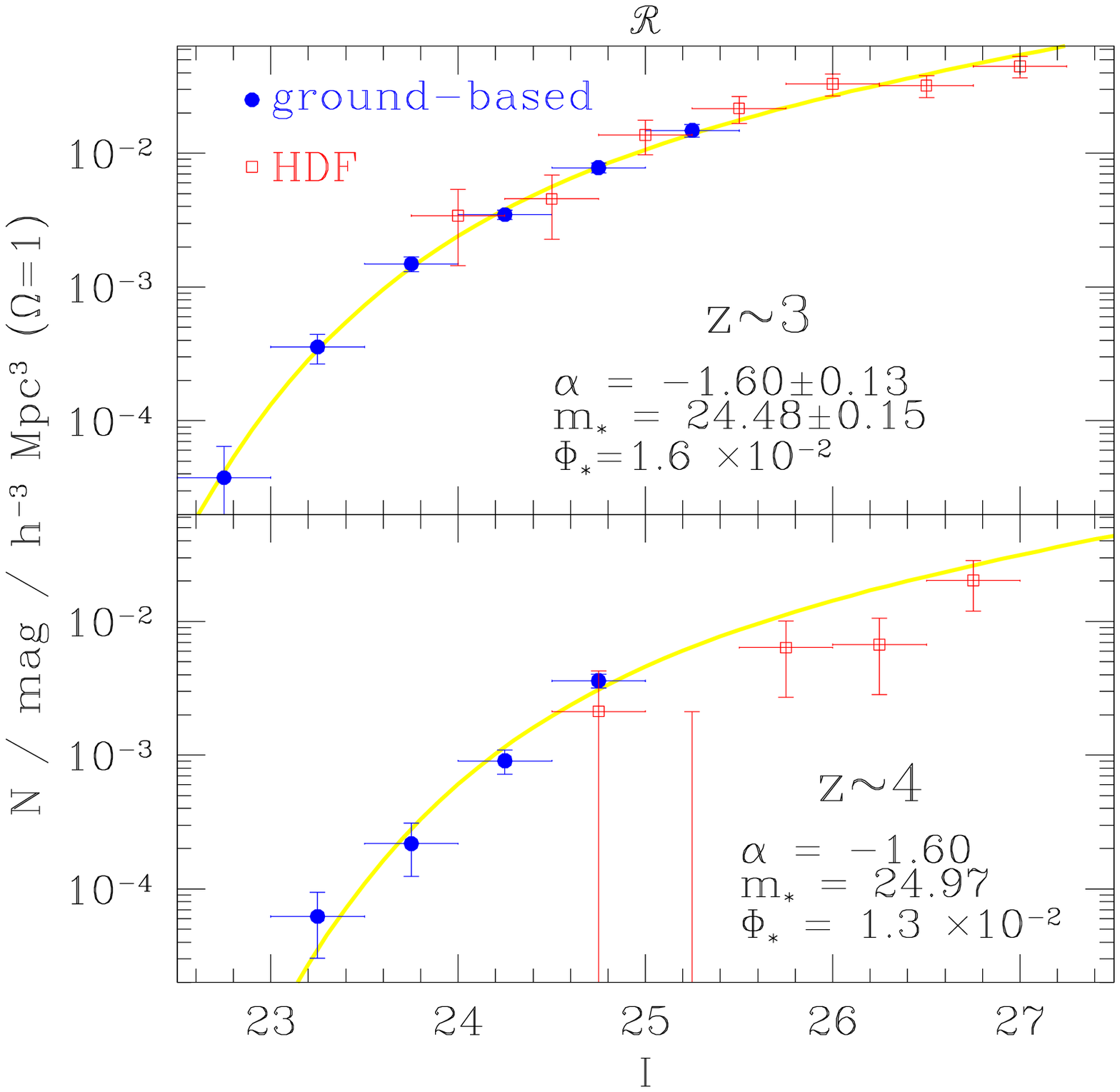}
\caption{Composite ground-based and HDF luminosity distributions for
$z \sim 3$ (top) and $z\sim 4$ (bottom). The HDF points account for
our new assessment of effective survey volumes. The results of a Schechter
(1976) function fit to the $z \sim 3$ luminosity distribution is shown
in the top panel. 
In the bottom panel,
the curve shown is the $z \sim 3$ luminosity function shifted by
0.49 magnitudes (the relative distance modulus between $z=3.04$ and
$z=4.14$ for an Einstein-de Sitter model) with the normalization 
adjusted to 80\% of the $z \sim 3$ value (cf. Table 5, Figure 5). 
The error bars for the
HDF--based points in both panels implicitly assume no clustering of the
LBGs (whereas the ground--based points do include this effect) and
the ``true'' error bars are probably significantly larger.}
\end{figure}
\begin{figure}
\figurenum{9}
\plotone{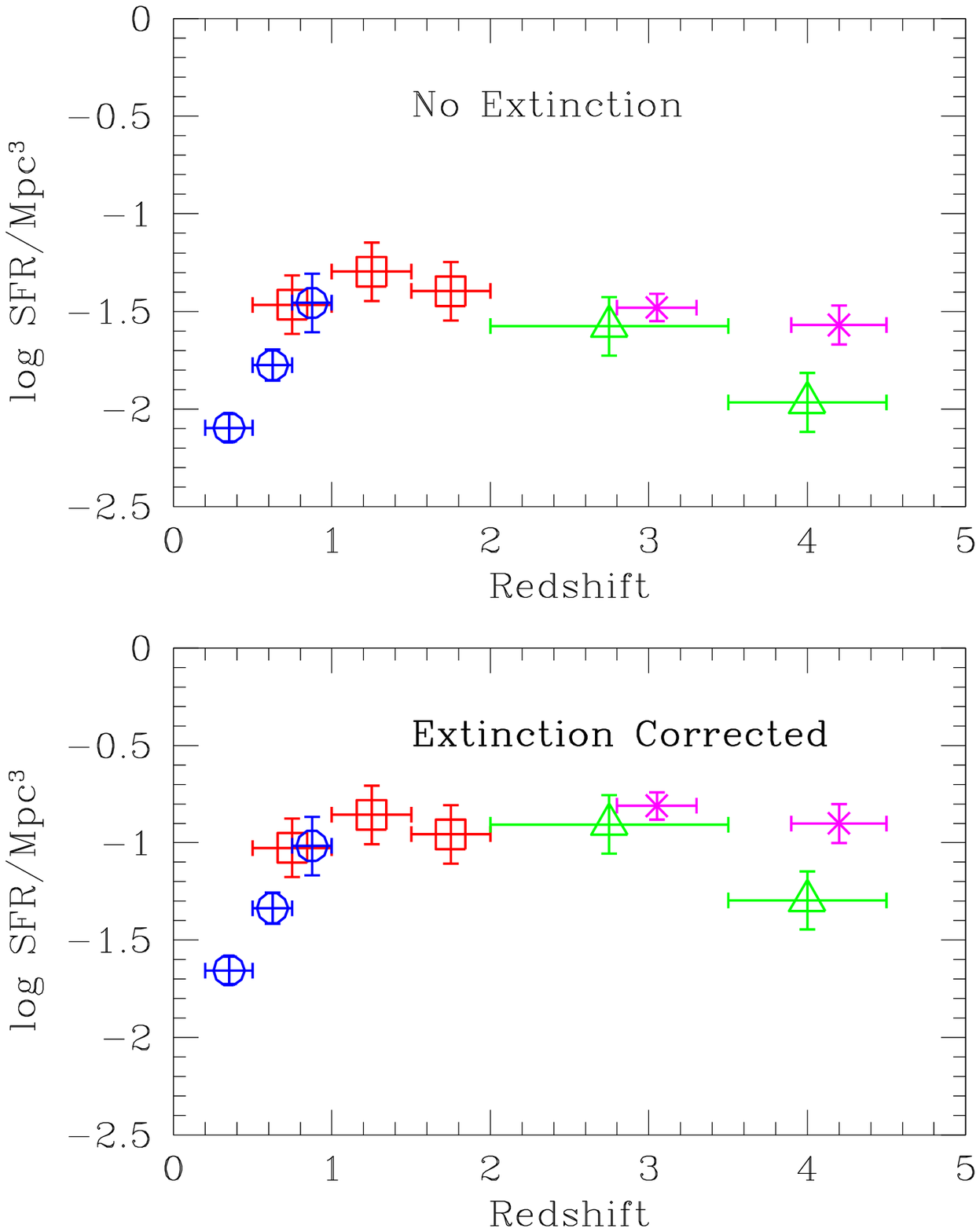}
\caption{The UV luminosity density as a function of redshift, following Madau \et 1996
(also using $H_0=50$ \kms Mpc$^{-1}$ and $q_0=0.5$, for consistency). The 
different points come from Lilly \et (1996) [circles], Connolly \et (1997) [squares],
and Madau \et 1997 (triangles). The new points from this work are shown as crosses.
See text for details.) 
}
\end{figure}
\end{document}